\documentclass[sts]{imsart}

\usepackage{amsthm,amsmath,amsfonts,amssymb}
\RequirePackage[authoryear]{natbib}

\usepackage{mathtools}
\usepackage{graphicx}
\usepackage{caption}
\usepackage{subcaption}
\usepackage{amsfonts}
\usepackage{float}
\usepackage{array}
\usepackage{multirow}
\usepackage{bm}
\usepackage[aboveskip=1pt,labelfont=bf,labelsep=period,justification=raggedright,singlelinecheck=off]{caption}
\usepackage{changepage}

\usepackage{textcomp,marvosym}

\usepackage{nameref,hyperref}

\usepackage[right]{lineno}
\usepackage[shortlabels]{enumitem}

\usepackage{microtype}

\usepackage{adjustbox} 
\usepackage[table]{xcolor}

\usepackage{array}
\startlocaldefs
\newcolumntype{+}{!{\vrule width 2pt}}

\newlength\savedwidth

\newcommand{\xhdr}[1]{\vspace{0.1mm}\noindent{{\bf #1.}}}
\usepackage{array}
\newcolumntype{x}[1]{>{\centering\arraybackslash\hspace{0pt}}p{#1}
}
\newcolumntype{L}[1]{>{\raggedright\let\newline\\\arraybackslash\hspace{0pt}}m{#1}}
\newcolumntype{C}[1]{>{\centering\let\newline\\\arraybackslash\hspace{0pt}}m{#1}}
\newcolumntype{R}[1]{>{\raggedleft\let\newline\\\arraybackslash\hspace{0pt}}m{#1}}

\newcommand{\indep}{\perp \!\!\! \perp}
\endlocaldefs
\usepackage{lastpage,fancyhdr,graphicx}
\usepackage{epstopdf}

\begin{document}

\begin{frontmatter}

\title{Statistical Modeling for Practical Pooled Testing During the COVID-19 Pandemic}
\runtitle{Statistical Models for Practical Pooled Testing}

\begin{aug}
\author[A]{\fnms{Saskia} \snm{Comess}\ead[label=e1]{saskiaco@stanford.edu}},
\author[B]{\fnms{Hannah} \snm{Wang}\ead[label=e2]{wangh1@stanford.edu}}, 
\author[C]{\fnms{Susan} \snm{Holmes}\ead[label=e3]{susan@stat.stanford.edu}}
\and
\author[D]{\fnms{Claire} \snm{Donnat}\ead[label=e4]{cdonnat@uchicago.edu}}


\address[A]{Saskia Comess is a PhD student in the Emmett Interdisciplinary Program in Environment and Resources at Stanford University \printead{e1}.}

\address[B]{Hannah Wang is a resident physician in the Department of Anatomic and Clinical Pathology at Stanford University School of Medicine \printead{e2}.}

\address[C]{Susan Holmes is a Professor in the Department of Statistics at Stanford University \printead{e3}.}

\address[D]{Claire Donnat is an Assistant Professor in the Department of Statistics at The University of Chicago \printead{e4}.}
\end{aug}

\begin{abstract}

Pooled testing offers an efficient solution to the unprecedented testing demands of the COVID-19 pandemic, although with potentially lower sensitivity and increased costs to implementation in some settings. Assessments of this trade-off typically assume pooled specimens are independent and identically distributed. Yet, in the context of COVID-19, these assumptions are often violated: testing done on networks (housemates, spouses, co-workers) captures correlated individuals, while infection risk varies substantially across time, place and individuals. Neglecting dependencies and heterogeneity may bias established optimality grids and induce a sub-optimal implementation of the procedure.\\
\indent As a lesson learned from this pandemic, this paper highlights the necessity of integrating field sampling information with statistical modeling to efficiently optimize pooled testing. Using real data, we show that (a) greater gains can be achieved at low logistical cost by exploiting natural correlations (non-independence) between samples —allowing improvements in sensitivity and efficiency of up to 30\% and 90\% respectively; and (b) these gains are robust despite substantial heterogeneity across pools (non-identical). Our modeling results complement and extend the observations of \citeauthor{Barak2021} who report an empirical sensitivity well beyond expectations. Finally, we provide an interactive tool for selecting an optimal pool size using contextual information\footnote{The corresponding shiny-app interface is available at: \url{https://homecovidtests.shinyapps.io/Group-testing/}}.

\end{abstract}


\end{frontmatter}

\pagebreak
\section*{Introduction}\label{sec:intro}

With an estimated 16\% of SARS-CoV-2 cases being asymptomatic and 50\% of detections occurring prior to symptom onset (\cite{he2020proportion,oran2020prevalence,pollock2020asymptomatic}), widespread surveillance testing plays a crucial role in monitoring and controlling the spread of SARS-CoV-2  (\cite{Larremore2020TestScreening.,gandhi2020asymptomatic,doi:10.1056/NEJMp2025631,dhillon2015ebola,nouvellet2015role,rannan2021increased}).  Yet in practice, the inherent logistical costs of widespread testing have severely limited its deployment at scale. Throughout the pandemic, testing needs have outstripped availability: in November 2020, the United States fell short of its COVID-19 testing objective by 48\%, performing a daily average of 1,193,000 tests out of the 2.3 million set as a minimum target (\cite{nyt}). Testing shortages continue to persist globally, as reported during the Spring 2021 surge in India (\cite{npr_india}) or as projected for the  "third wave" throughout Africa in Summer 2021  (\cite{bbc_africa}). 
The unprecedented surge in testing demand  has also strained the broader laboratory supply chain; from September 2020 through January 2021 in the U.S., shortages of testing materials (e.g. reagents, consumables, etc) significantly impacted day-to-day testing for both COVID-19 and other infectious diseases (\cite{asm_testing}). 
Consequently, despite rising vaccination rates in some parts of the world (at the time of writing), the threats of new variants, waning immunity, and localized outbreaks  
make the deployment of robust and continued large scale testing a priority.

In this context, pooled (or group) testing procedures have generated increasing interest during the pandemic (\cite{Abdalhamid2020AssessmentResources}). First proposed by \cite{Dorfman1943} to screen soldiers for syphilis, the simplest form of pooled testing is a two-stage hierarchical procedure in which multiple laboratory specimens are first combined and tested, and samples from positive pools are subsequently individually re-tested. Since then, pooled testing has been successfully employed in a number of applications,  ranging from the testing for low prevalence diseases (including HIV, chlamydia, and gonorrhea (\cite{McMahan2012InformativeScreening,wein1996pooled,tu1995informativeness,gaydos2005nucleic}), to the detection of genetically modified organisms in crops (\cite{yamamura2007estimation}).

Despite limited prior use of pooled testing for widespread epidemic management, the American Food and Drug Administration (FDA) approved its use for certain SARS-CoV-2 diagnostic tests in Summer 2020, with some restrictions. According to these guidelines, the sensitivity of the pooled procedure should be maintained above the threshold of 85\% (\cite{FDA2020CoronavirusTesting}). 
This desideratum must also be weighed against logistical feasibility of implementing pooled testing --- a tension recently described by the College of American Pathologists\footnote{\url{https://www.cap.org/covid-19/pooled-testing-guidance-from-cap-microbiology-committee}} and which can be summarized along the following two axes:\\
\indent \textbf{\textit{{(a) Logistics.}}} Prior to the pandemic, a substantial body of literature already considered better pooling designs ---  either by grouping specimens according to a set of covariates (\cite{McMahan2012InformativeScreening,chen2009group,bilder2010informative}), or by placing the samples into an array matrix (\cite{mcmahan2012two}) to allow the immediate identification of contaminated specimens without individual retesting. While these procedures can yield impressive efficiency gains from a purely statistical perspective, they simultaneously introduce more room for errors in specimen handling: if performed manually, specimen pooling can increase risk of specimen confusion and cross-contamination while increasing lab handling times. Automated robots become  essential for aliquoting and attributing samples to pools following complex optimal designs. Thus, while mathematically optimal, these solutions are often difficult to implement without state-of-the-art (and often expensive) equipment.  \\
\indent \textbf{\textit{{(b) Context-dependent efficiency and sensitivity.}}} The sensitivity and efficiency (number of tests per sample) of pooled testing are known to be functions of the pool size and disease prevalence. The latter determines the probability that a pool contains at least one positive individual, and therefore, that all individuals in the pool require re-testing (\cite{Gastwirth2000The2}): larger pool sizes in low prevalence regimes ensure that fewer tests have to be carried out. Concurrently, pooling dilutes the amount of viral genetic material present in positive samples, thereby potentially reducing the sensitivity of the procedure. Previous studies have investigated the sensitivity of pooled testing under different prevalence levels in order to develop coarse recommendations for selecting an appropriate pool size at a given prevalence level (\cite{Kim2007ComparisonError,McMahan2012InformativeScreening}). Such studies typically assume independently and identically distributed (i.i.d) samples when estimating the appropriate pool size. However, such i.i.d assumptions may not be reasonable for SARS-CoV-2 given widespread community transmission and specimen collection procedures that capture highly connected networks --- a phenomenon virtually unique to an epidemic scenario, and irrelevant to the low-prevalence-disease monitoring that pooled testing has historically been used for. 

In fact, in many SARS-CoV-2 collection scenarios, infections and positive specimens are clustered, such as when testing students in dorms, coworkers in an office,  individuals in households, or classrooms of children on a weekly basis (\cite{wapo_schools, Adam2020ClusteringKong,Zhang2020FamilialAsymptomatic}). Transmission rates in these networks may be much higher than in less connected networks (Table \ref{table:tau_dist} in Appendix \ref{sec:appendix_sim}). By way of illustration, household transmission rates are estimated to vary from 4\% to 55\% (\cite{Koh2020})--- highlighting  transmission rates within close communities greater than the overall prevalence by orders of magnitude, but with significant uncertainty and heterogeneity. This leads us to a first lesson that we draw from our analysis: \\
\xhdr{Lesson 1} \textit{ The unprecedented data collection and sampling processes observed during the pandemic have severely compromised the validity of classical statistical pipelines for the analysis of data --- thereby  leading to an inaccurate evaluation of pooled testing and  potentially suboptimal deployment of this method. Statistical modeling  is key to rapidly and efficiently re-adapting existing procedures to this novel setting, but its relevance is contingent on being able to bridge   the gap between statistics-based optimality --- which strives to make the greatest efficiency gains --- and field-based optimality -- which is informed by practical constraints and logistical considerations. } 

To adapt to this novel situation, several studies have begun investigating the impact of deviations from the i.i.d setting on the sensitivity and efficiency of pooled testing. To account for \textit{non-independence},  \cite{Rewley2020SpecimenStudy} simulated correlations between consecutive persons in a testing queue, assuming an additive increased chance of a positive test given the previous person in line was positive. The simulations suggested that as the likelihood of clustered infections increased, pooling efficiency also increased, even with rising prevalence. Other simulation studies have similarly concluded that the Dorfman two-stage procedure is optimal when testing is performed on clusters of correlated individuals  (\cite{Deckert2020SimulationTesting, Lin2020PositivelyCosts}). So far, many of these studies have necessarily relied on simulating simplified settings, with arbitrary parameterisations and distributions and ignoring variability across pools. Few simulation studies have attempted to capture deviations from the \textit{identically distributed} hypothesis, in part because there is minimal practical applicability for incorporating information on individual level covariates relevant to infection risk. 

\indent Taking an experimental data-driven approach, \cite{Barak2021} examined Dorfman-based pooled testing on over 130,000 SARS-CoV-2 samples. Pool size was adaptively chosen based on predicted prevalence levels in the community. They found that the rate of positives in pooled samples was best predicted by sorting samples into batches according to their source (such as by colleges, nursing homes or health care personnel) and also incorporating epidemiological information about the probability of infection in these different sources. Overall, they observed that pooled testing performance exceeded expectations both in terms of efficiency and sensitivity, which they attributed solely to the fact that there is a nonrandom distribution of positive samples in pools. Thus, real-world data supports the push to develop easily adaptable pooled testing strategies that exploit the non-i.i.d nature of samples.  Due to rapidly changing SARS-CoV-2 prevalence, laboratories require practical tools that allow them to adapt their procedures to the context and populations they treat.


\xhdr{{Objectives}} 
We have two primary objectives:
 \begin{enumerate}[nosep]
    \item \textit{Show how a simple pooling method and accounting for correlated specimens in statistical modeling can yield unexpectedly efficient solutions}. To this end, we provide a straight-forward model that measures the efficiency of pooled testing under correlations, as well as formalizes and extends the lessons from \cite{Barak2021} in practical pooled sampling. 
    \item \textit{Investigate and produce actionable recommendations that are ready for deployment during the COVID-19 pandemic}. For this reason, we focus on the Dorfman two-stage procedure rather than more mathematically optimal, but unscalable, pooling procedures. 
\end{enumerate} 
\indent We organize our discussion around three main lessons that our investigation of pooled testing for SARS-CoV-2 samples has taught us: (1) the importance of leveraging setting-specific information to optimize testing, (2) the necessity of evaluating efficiency through a set of several, practically meaningful measures, and (3) the importance of modeling the impact of uncertainty and/or heterogeneity. We show that pooled testing can efficiently identify infectious individuals despite natural deviations from i.i.d hypotheses in the specimen collection process, with little detrimental effect on the accuracy of the procedure; 
Gains in sensitivity and efficiency can in fact be as much as 30\% and 90\% respectively compared to i.i.d settings.  In contrast to existing studies, our modeling of correlations (i) is focused on understanding pooled sampling's heightened sensitivity through its effect on the viral load of the sample (measured as the cycle threshold ($C_t$) value), (ii) is informed by real data and practical constraints and above all, (iii) allows for the simultaneous consideration of non-independence and population heterogeneity.



\section{Leveraging correlations and the hitchhiker effect}\label{sec:maths_framework}

This section focuses on a three-fold approach to understanding the impact of deviations from the i.i.d hypotheses on the $C_t$ value: (1) We introduce a model that accounts for positive correlations between samples; (2) We expand upon the observation by \cite{Barak2021} that pooled testing achieves improved sensitivity of pooled testing due to the "hitchhiker effect" --- a phenomenon whereby the detection of weakly positive tests is improved by borrowing strength from strongly positive tests --- by providing a mathematical framework to quantify this effect; and (3) We use this framework to show how positive correlations between samples can further improve the sensitivity of the pooling procedure.  

\subsection{A Network Model for Pooled Specimens}


\xhdr{Modeling Non-independence (Network Effect)} Consider that there are $n$ samples in each pool. We assume that all the specimens are sampled from a network (co-workers, classroom, household, etc.) modeled by a fully connected graph on $n$ nodes with edges indicating potential transmission between a given node and its neighbors. We denote by $\tau$ the network transmission probability between individuals in the pool (that is, the probability that an infected member infects another subject in the network), so that each edge $e_{ij} = \tau$ represents the probability that node $i$, if infected, transmits the disease to node $j$; in epidemiology, this parameter is commonly referred to as the secondary attack rate (SAR).  The community transmission probability, or equivalently, the prevalence in the population, is denoted by $\pi$. Let us denote as $Y_i$ the indicator variable that specimen $i$ is infected ($Y_i = 1$ if individual $i$ infected, $0$ otherwise). Since transmission can occur within the network or in the community, we decompose $Y_i$ as follows: (1) Let $T_i^{(cmty)}$ be the random variable indicating infection of individual $i$ from outside the group (community transmission), and (2) Let $T_i^{(ntw)}$ indicate infection from within the group (network transmission). The existence of a network effect is captured by writing the infectious status of individual $i$ as the sum: 
$$  Y_i = T_i^{(\text{cmty})} + (1-T_i^{(\text{cmty})})T_i^{(\text{ntw})} \quad \text{with } T_i \in \{0,1\},\quad  T_i^{(\text{cmty})} \indep  T_i^{(\text{ntw})}  \hspace{1cm}(M_1),$$
so that $$ \mathbb{P}[Y_i=1] = \mathbb{P}[ \{T_i^{(\text{cmty})}=1\} \cup \{T_i^{(\text{net})}=1\}].$$ 

Network and community transmissions are themselves modeled as Bernoulli variables and tied to the SAR $\tau$ and prevalence $\pi$ through the relations: 
$$\mathbb{P}[T_1^{(\text{cmty})} = 1 ] =\pi  \hspace{1cm} \text{ and }  \hspace{1cm} \mathbb{P}[T_1^{(\text{ntw})} = 1 | Y_2, \cdots, Y_n] = 1 - (1-\tau)^{\sum_{i=2}^n T_i^{(\text{cmty})}}.$$ In other words, network transmission is modeled as independent Bernoulli($\tau$) variables across edges, so that the probability that this transmission route fails is the product of the probability of failure across each edges: $\mathbb{P}[T_1^{(\text{ntw})} = 0 | Y_2, \cdots, Y_n]  = (1-\tau)^{\sum_{i=2}^n T_i^{(\text{cmty})}}$.  A given pool can have a total of $K \le n$ positive samples, $k$ of which are infected from the community and $K-k$ of which are infected from within the network.\\

\begin{figure}[h]
    \centering
    \includegraphics[width=10cm]{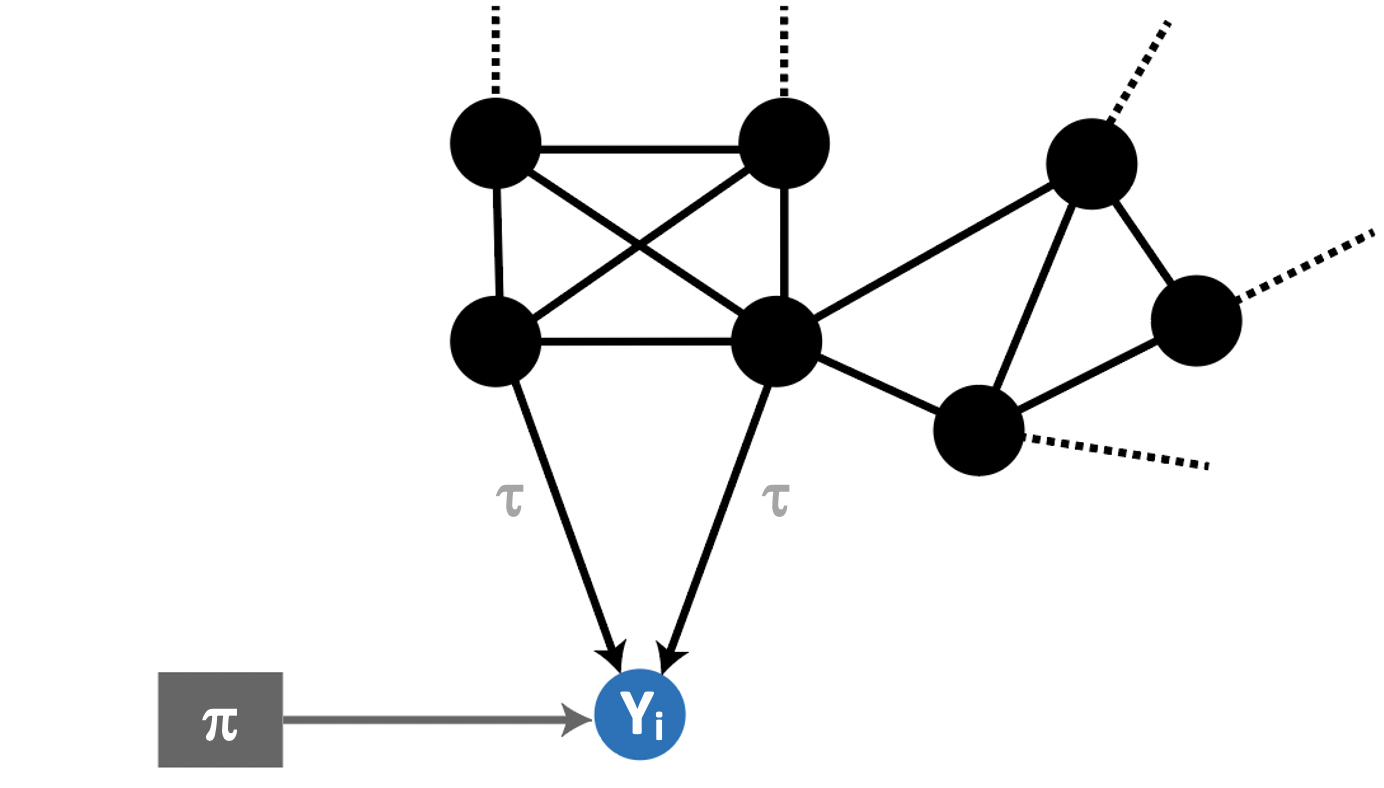}
    \caption{Representation of the network effect on the infectious status for any given node. This highlights the higher risk of sample $i$ testing positive as soon as another sample in the network is also positive.}
    \label{fig:net_effect}
\end{figure}

\xhdr{Modeling Non-identical Distributions (Heterogeneous Infection Probabilities)} Individuals within a network are exposed to various levels community transmission risk depending on a number of covariates, including age, profession, and lifestyle. This risk also varies considerably with time and epidemic kinetics (new variants, vaccination levels, etc.). At the granular level, this can be captured by introducing node covariates $X_i$'s and allowing heterogeneous community infection rates:
$$ \forall i, \qquad \mathbb{P}[T_i^{(\text{cmty})} = 1] = \pi  + f(X_i)  + \epsilon_i$$
where $\pi$ is the general community prevalence level, $f(X_i)$ reflects deviations from this baseline level depending on a set of covariates (e.g. profession, lifestyle), and $\epsilon_i$ is a noise term capturing the stochasticity of the prevalence (e.g. temporal effects) and/or potential subject effects. 
Several studies have investigated using subject-level covariates to inform the risk function $f$, by classifying individuals either as high/low risk (\cite{McMahan2012InformativeScreening,bilder2019informative, donnat2020bayesian}), or to inform retesting (\cite{bilder2012pooled}).
However, in the context of the COVID-19 pandemic, and due to the necessary volume and frequency of testing, introducing individual subject covariates yields impractical solutions: the collection of subject-level covariates and dispatching of samples in pools according to their risk slows down the procedure and yields intricate pooling designs which are not feasible at scale. As such, we  propose simply leveraging the specimen collection process to assume similar behaviors and covariates across pooled specimens ($f(X_i) \approx f(X_j) \quad  \forall i, j$) so that pools on any given day can be considered homogeneous, and that random effects --- due for instance to temporal variations, different variants, etc --- are at the pool (instead of the individual subject) level: 
$$ \forall i, \qquad \mathbb{P}[T_i^{(\text{cmty})} = 1]  = \pi  + \epsilon_{\text{pool}}.$$

\xhdr{Discussion} This simple model allows us to capture a variety of situations. In particular, (i) the value of the SAR $\tau$ drives the balance between community and network transmission: the higher the value of $\tau$, the more likely that any community transmission will yield more than one secondary infection in the pool. Conversely, for $\tau=0$, there is no network transmissions and the samples are independent.  (ii) Assuming this notation, the standard i.i.d case studied in the literature is a homogeneous, fixed effect model and assumes $f=0$ and $\epsilon=0$  (so that $\forall i, \quad \pi_i = \pi$) and $\tau=0$. We refer to the uncorrelated, i.i.d scenario with prevalence $\pi$ and pool size $n$ as our null model $H_0(\pi, n)$, and to the correlated, heterogeneous one as the alternative $H_a(\pi + f,  n, \tau)$. (iii) This model allows us to study the law of the number of infected samples in the pool. As we will describe in subsequent subsections, the shift in probability distribution leads to improved performance of the pooling procedure. In Table~\ref{tab:properties}, we summarize the properties of the distribution of number of positive samples per pool under both models. 

\begin{table}
    \caption{Comparison of the properties of the number of infected samples per pool (size=$n$), with, and without correlation. All the derivations are provided in Appendix \ref{sec:appendix_comp}. Note, in particular, the shift in the expected number of samples as a function of the network  transmission rate (or SAR), $\tau$.}
    \label{tab:properties}
    \centering
    \resizebox{\textwidth}{!}{\begin{tabular}{{@{}lll@{}}} 
    \hline
       Quantity  & $H_0(\pi, n)$ &  $H_a(\pi,  n, \tau)$\\  
    \hline
       Distribution  & $\sum_{i=1}^n Y_i \sim \text{Binomial}( n, \pi ) $  &  $\sum_{i=1}^n Y_i$ + Binomial$(\sum_{i=1}^n Y_i, \tau )$ ;  with $\sum_{i=1}^n Y_i \sim \text{Binomial}( n, \pi ) $  \\ [10pt]
       
       $\mathbb{P}[Y_i=1]$  &  $\pi$  & $1-(1-\pi)(1-\pi\tau)^{n-1}$ \\ [10pt]
       
       Correlation $c(Y_i, Y_j)$  & 0 & $\frac{(1-\pi)^2\Big[(1-\pi + (1-\tau)^2 \pi  )^{n-2}-  (1-\pi\tau)^{2n-2}\Big]}{{(1-(1-\pi)(1-\pi\tau)^{n-1})((1-\pi)(1-\pi\tau)^{n-1})}}$\\ 
       & & $=(1+\frac{1-\tau}{1+(n-1)\tau})\tau + o(\tau)$ \\ [10pt]
       
       Average number of & $n\pi$  & $n\pi$ + $\sum_{k=1}^{n-1} {n \choose k} \pi^k (1-\pi)^{n-k} (1-  (1-\tau)^k) (n-k)$ \\
       positive samples per pool & & $= n\pi ( 1 + (n-1)\tau) + o(n\pi)$\\[10pt]
       
       Average number of & $\frac{n\pi}{1-(1-\pi)^n}$  & $1 + \sum_{k=1}^{n-1} {n \choose k} \frac{\pi^k (1-\pi)^{n-k}}{1-(1-\pi)^n} (1-  (1-\tau)^k) (n-k)$ \\
       positive samples  & $\approx 1$ & $\approx( 1 + (n-1)\tau)$ \\  
          per contaminated pool    & & \\ [10pt]
          
        $\mathbb{P}[\sum_{i=1}^n Y_i=0] $ & $(1-\pi)^{n}$ &  $(1-\pi)^{n}$  \\ [10pt]
        
        $\mathbb{P}[\sum_{i=1}^n Y_i=k], k>0$&  ${n \choose k} \pi^k (1-\pi)^{n-k}$ & $ \sum_{j=1}^k {n \choose j}  \pi^j (1-\pi)^{n-j} {n - j \choose k-j} (1-  (1-\tau)^j)^{k-j} (1-\tau)^{ j (n-k)}$ \\[.1cm] \hline 
    \end{tabular}}

\end{table}

Since (as in Table~\ref{tab:properties}), the probability of the number of positive samples per pool is a complex polynomial function of $\tau$, we propose an approximation when the prevalence is small. This allows us to gain greater insight into the intricate interplay between $\tau$ and $\pi$, without hindering the utility of the analysis since pooled sampling is predominantly deployed in low prevalence settings. More precisely, we make the following assumption:
\begin{quote}
\textbf{Assumption 1- Low Prevalence}: We assume that the prevalence is such that $\pi n \leq 0.10$. To put this number into context, this scenario is aligned with situations observed in Summer 2020 or Spring 2021 in Europe and the U.S.: in early June 2021 for instance, the reported prevalence of COVID-19 in the United Kingdom was estimated around 0.70\%, which would allow us to look at pool sizes of up to 15, or to sizes of up to 50 for low prevalence levels under 0.2\% observed in some parts of the world (such as for instance Israel \footnote{Data from \href{https://ourworldindata.org/covid-cases}{Our World in Data} (percentage of positive samples per test). For the sake of completeness, a chart is provided in Figure~\ref{fig:comp} in Appendix \ref{sec:appendix_complementary}.}) in late June 2021 . This is a convenient threshold that allow us to simplify the analysis while still providing insight into the interplay between community and network transmission, as highlighted by the two following observations.
\end{quote}

\noindent \textbf{Observation 1: } Under Assumption 1 ($\pi n \leq 0.10$), the probability that under $H_0$, there are two or more infected samples in the pool is less than $.01$.
This follows from the following simplification:
\begin{equation*}
    \begin{split}
        \mathbb{P}_{H_0}[\sum_{i=1}^n Y_i > 1] &= 1- (1-\pi)^{n} - n\pi(1-\pi)^{n-1} = 1 - (1-\pi)^{n-1} ( 1 + (n-1)\pi)\\
        &\leq  1 -  (1-(n-1)\pi + \frac{(n-1)(n-2)}{2}\pi^2 - \frac{(n-1)(n-2)(n-3)}{6}\pi^3 + o(n^3\pi^3) )  \\
        &\times( 1 + (n-1)\pi)\\
        &\leq \frac{n(n-1)}{2} \pi^2 -n \frac{(n-1)(n-2)}{3} \pi^3 +  o(n^3\pi^3)\\
        &\leq 0.005. 
    \end{split}
\end{equation*}
Similarly, in a pool with at least one infected sample, we can show that:
$$     \mathbb{P}[S \geq 2| S \geq 1 ] \leq  \frac{n}{2}\pi \leq 0.05.$$
(see Appendix \ref{sec:appendix_comp} Eq.\ref{eq:eq} for details). This means that with 95\% confidence, an infected pool contains only a single infected sample. This fact will be useful to simplify our computations in the subsequent paragraphs.

\noindent \textbf{Observation 2: }  Concurrently, in this scenario, the number of positive samples per pool \textbf{with correlation} is: 
$\mathbb{E}_{H_a}[\sum_{i=1}^n Y_i]= n\pi + n(1-\pi)  (1- (1-\tau\pi)^{n-1}) $ while the probability that there are more than 1 infected samples in the pool is:
\begin{equation*}
    \begin{split}
        \mathbb{P}_{H_a}[\sum_{i=1}^n Y_i > 1] &= 1- (1-\pi)^{n} - n\pi(1-\pi)^{n-1} (1-\tau)^{n-1}\\ 
        &= \mathbb{P}_{H_0}[\sum_{i=1}^n Y_i > 1]   + \big( 1- (1-\tau)^{n-1} \big) n\pi(1-\pi)^{n-1}  \\ 
        &\geq  \mathbb{P}_{H_0}[\sum_{i=1}^n Y_i > 1] \Big( 1 +  \frac{2(1-(1-\tau)^{n-1})}{(n-1)\pi}  - \frac{2((n+1)((1-\tau)^{n-1}  -1))}{3(n-1)} \Big)\\
        &\approx  200  (1 -(1-\tau)^{n-1}) \mathbb{P}_{H_0}[\sum_{i=1}^n Y_i > 1] \quad \text{ under Assumption 1.}\ 
    \end{split}
\end{equation*}
where the last line follows from the fact that the dominating term is in $\frac{1}{\pi} \geq 100$, whereas the others  are of order $\sim 1$ (see Appendix \ref{sec:appendix_comp}  Eq.~\ref{eq:proba_ha} for details).  This illustrates the striking difference in behavior between the distribution of positive samples with and without correlation: the probability of having more than two samples in a correlated pool increases rapidly, as the inverse of the prevalence. 

\begin{figure}[ht]
    \centering
    \includegraphics[width=\textwidth]{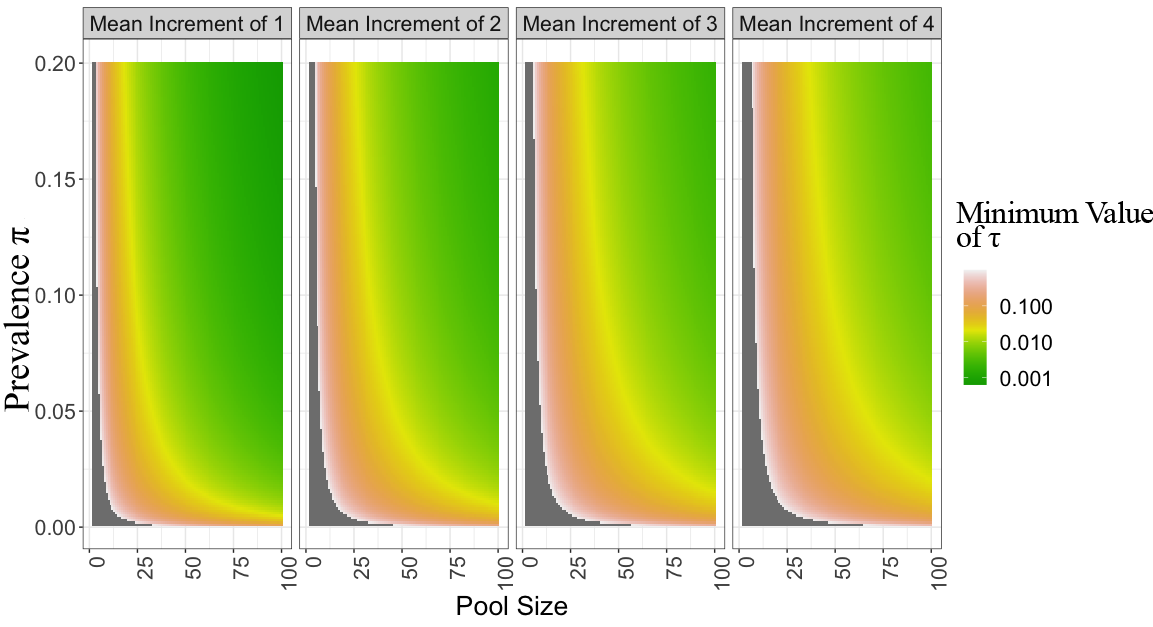}
    \caption{Minimum value of $\tau$ required to shift the expected number of infected samples in a pool by $k$ as a function of the pool size and the overall community prevalence $\pi$. NA values are highlighted in grey. We note that small values of $\tau$ (e.g, 0.01, here represented by the yellow color) can already yield an increase of 1 or more in the expected number of positive samples.}
    \label{fig:increments_tau}
\end{figure}

\noindent \textbf{Observation 3: } Conditional on the pool being positive, the expectation of number of positive samples per pool \textbf{with correlation} can be well approximated by: $\mathbb{E}_{H_a}[\sum_{i=1}^n Y_i] = 1+(n-1) \tau +  \frac{1}{2}(n-1)(1-\tau)((n-2) \tau +1) \pi + O(n^3\pi^3)$, where the remainder is bounded by $n\pi$ (proof in Appendix \ref{sec:appendix_comp})) --- thereby shifting the mode of the number of infected samples (conditional on  $\sum_{i=1}^n Y_i>1$) from 1 under $H_0$ (with 95\% confidence) to $ 1 + (n-1)\tau + 0.05 (n-2)(1-\tau)\tau \approx 1 + (n-1)\tau  $ with correlations. The number of positive samples in infected pools thus increases roughly linearly with the SAR. This further supports the observation by \cite{Barak2021} regarding the existence of positive correlations between samples: our relationship quantifies the effect and the strength of the interactions on the distribution of the number of positive samples per pool. This relationship is illustrated in  Figure \ref{fig:increments_tau}, where we show the minimum value of $\tau$ required to induce an increment of $k=1, \cdots, n-1$. We observe that very low values of $\tau$ can shift the distribution quite considerably: for a pool size of 70, in low prevalence regimes, a value of $\tau=0.001$ is sufficient to increase the expected number of samples in positive pools by 2. In short, in an i.i.d setting, the probability of observing strictly more than one sample is extremely low. Our simple model highlights the fact that with moderate correlations between samples, observing more than one sample becomes in fact highly probable --- whereas it is unlikely with high probability (0.95) under the i.i.d null.


\subsection{Modeling the Hitchhiker Effect}

\begin{figure}[ht]
    \includegraphics[width=\textwidth]{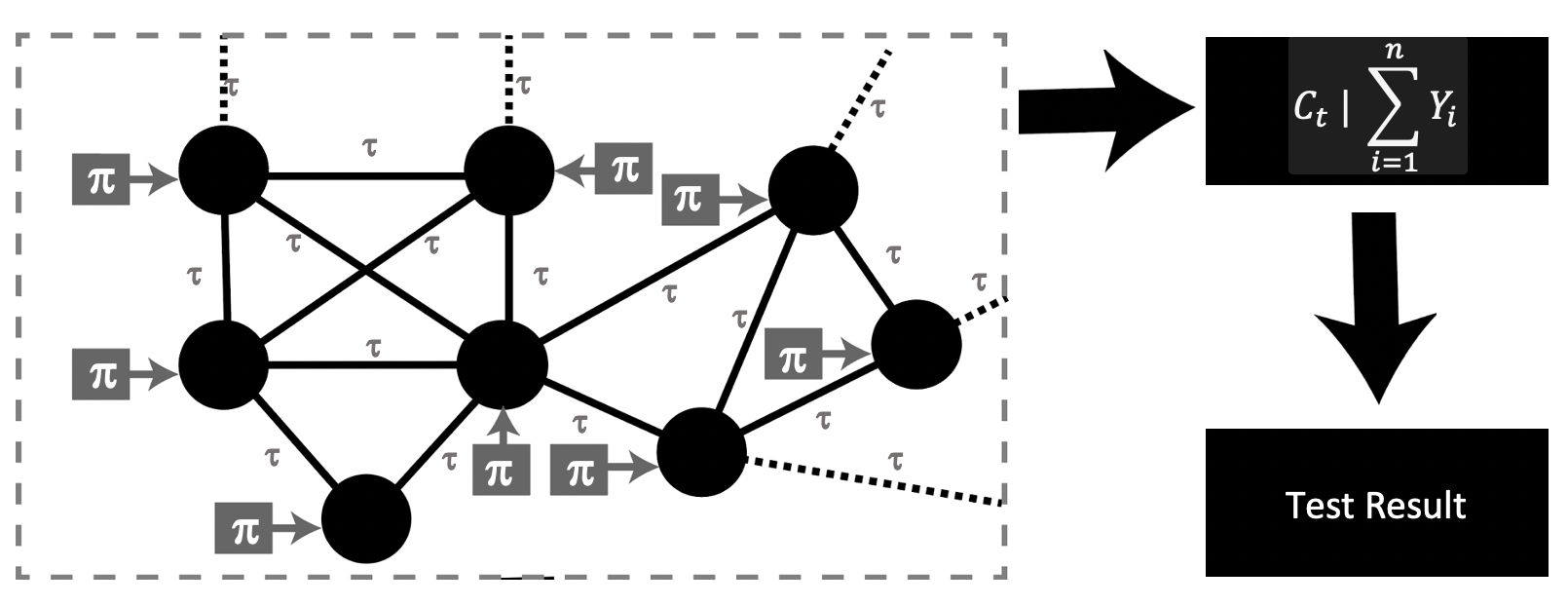}
    \caption{Graphical model for the procedure: the probability of a positive pooled test is a function of the number of infected samples (through community  and network  transmission) through the pooled $C_t$ value.}
    \label{fig:graph_model}
\end{figure}

Armed with the previous set of observations, we now turn to the quantification of the "hitchhiker effect" observed by \cite{Barak2021}--- a phenomenon whereby the detection of weakly positive samples borrows strength from strongly positive samples in the pool, thus increasing their chances of detection. This requires us to model the influence of the number of contaminated samples in the pool on the viral load present in the sample, as detailed in the generative model presented in Figure~\ref{fig:graph_model}.  This viral load is measured during testing by the \textit{cycle threshold} ($C_t$) of the pooled sample. The $C_t$ in reverse transcription polymerase chain reaction (RT-PCR) refers to the number of cycles needed to amplify viral RNA to reach a fixed background level $T$ of fluorescence at which the diagnostic result of the real-time PCR changes from negative (not detectable) to positive (detectable)\ (\cite{tom2020interpret}). Since at every cycle, the amount of viral RNA is (roughly) doubled, the $C_t$ value is thus an indirect measure of the viral load $n^{(0)}$ in the sample:

\begin{equation}
T = n^{(0)} 2^{C_t}  \implies C_t = \log_2(T/n^{(0)})
\label{eq:ct} 
\end{equation} 

\begin{figure}
    \begin{subfigure}[t]{0.49\textwidth}
        \centering
        \centering
    \includegraphics[width=\textwidth, height=4cm]{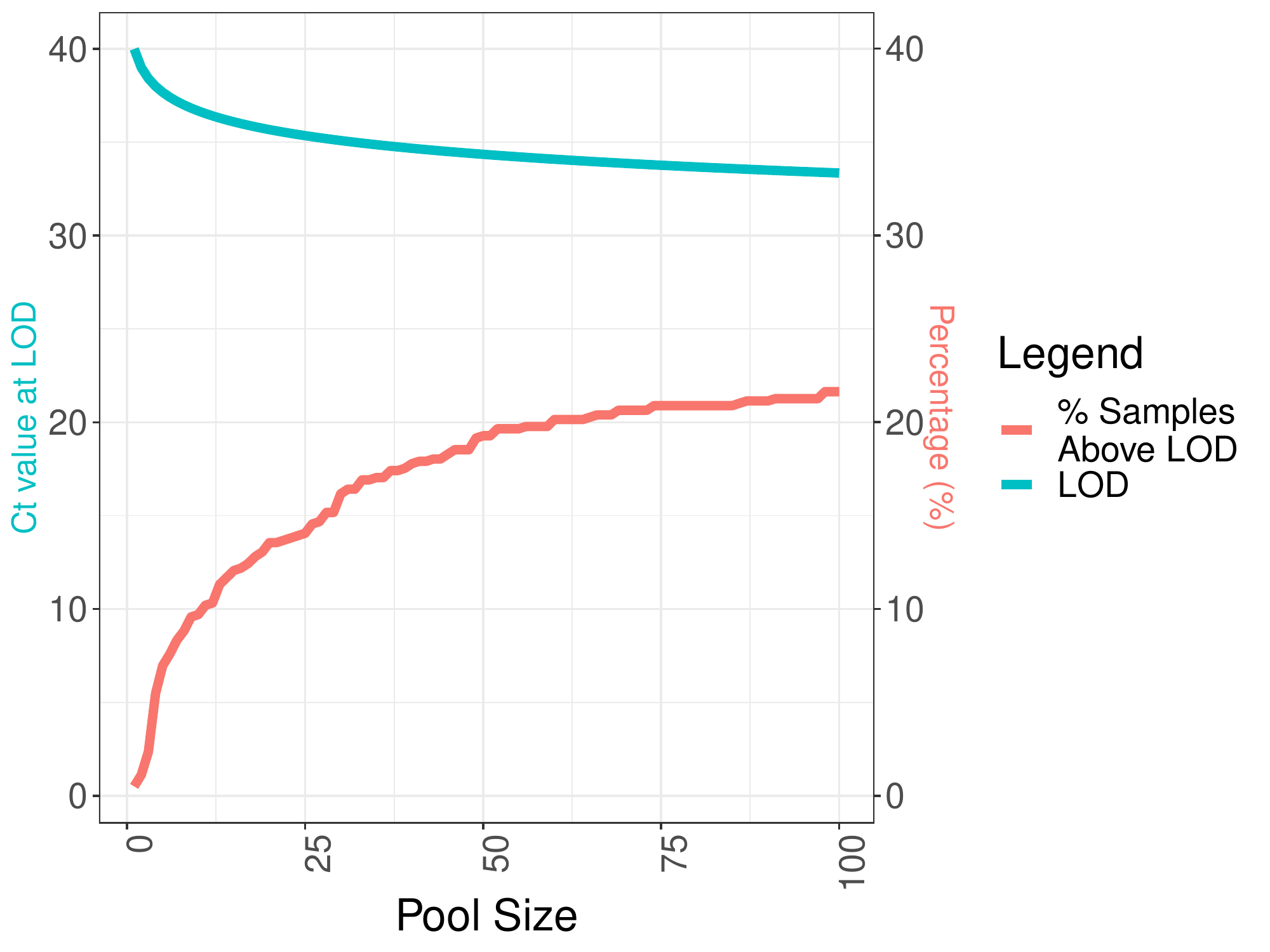}
    \caption{Pool size vs LOD and percentage of individual samples above the LOD}
    \label{fig:lod}
     \end{subfigure}  
     \begin{subfigure}[t]{0.5\textwidth}
    \centering
    \includegraphics[width=\textwidth, height=4cm]{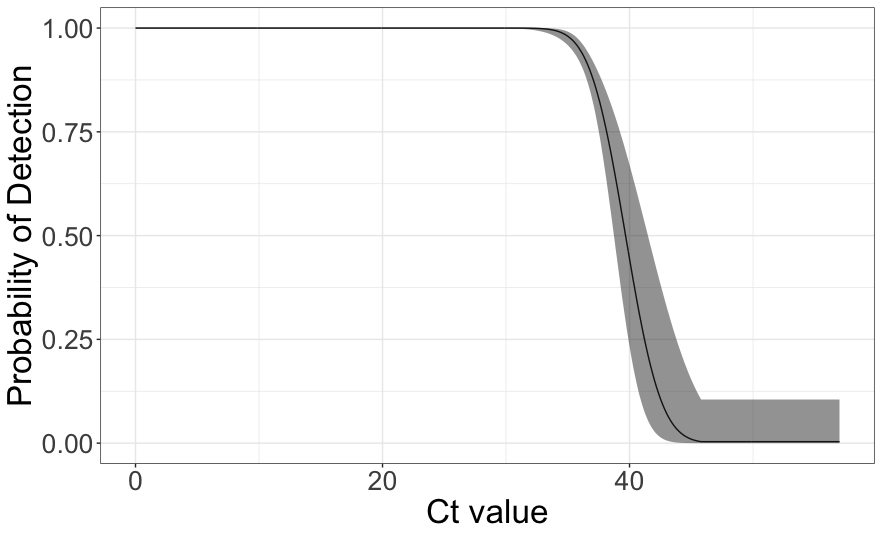}
    \caption{Sensitivity of the test as a function of the $C_t$ value}
    \label{fig:sens}
 \end{subfigure} 
     \caption{Probabilistic model for the metrics of interest and sensitivity as a function of the $C_t$ value}
\end{figure}

The amount of viral material determines the sensitivity of the test: according to the previous equation, the higher the amount of viral material, the lower the $\text{C}_t$ value and the higher the probability of detection (Figure~\ref{fig:sens}).

\xhdr{Effect of Dilution on $\text{C}_t$ Values}  Since by Eq. \ref{eq:ct}, $\text{C}_t = \log_2(T/n^{(0)})$, lowering the initial amount of viral material in the sample $n^{(0)}$ increases the $\text{C}_t$ value --- thus potentially lowering the initial viral load below the limit of detection and decreasing the sensitivity of the test. In fact, denoting as $\text{C}_{t}^{(i)}$ (respectively $n_i^{(0)}$) the individual samples' $\text{C}_t$ values (respectively  initial viral load), we can write the $\text{C}_t$ value for the dilution as:

\begin{equation} 
\begin{split}
    C^{\text{dilution}}_t &= \log_2(T) - \log_2(\frac{\sum_{i=1}^n n_{i}^{(0)}}{n})= \log_2(T) - \log_2(\frac{\sum_{i=1}^n T Y_i 2^{-C_{t}^{(i)}}}{n})\\
     &= - \log_2 ({\sum_{i=1}^n  Y_i2^{-C_{t}^{(i)}}}) +\log_2(n) 
    \end{split}
    \label{eq:dilution_ct}
\end{equation} 
 
\begin{figure}
    \begin{subfigure}[t]{0.49\textwidth}
        \centering
        \centering
    \includegraphics[width=\textwidth, height=5cm]{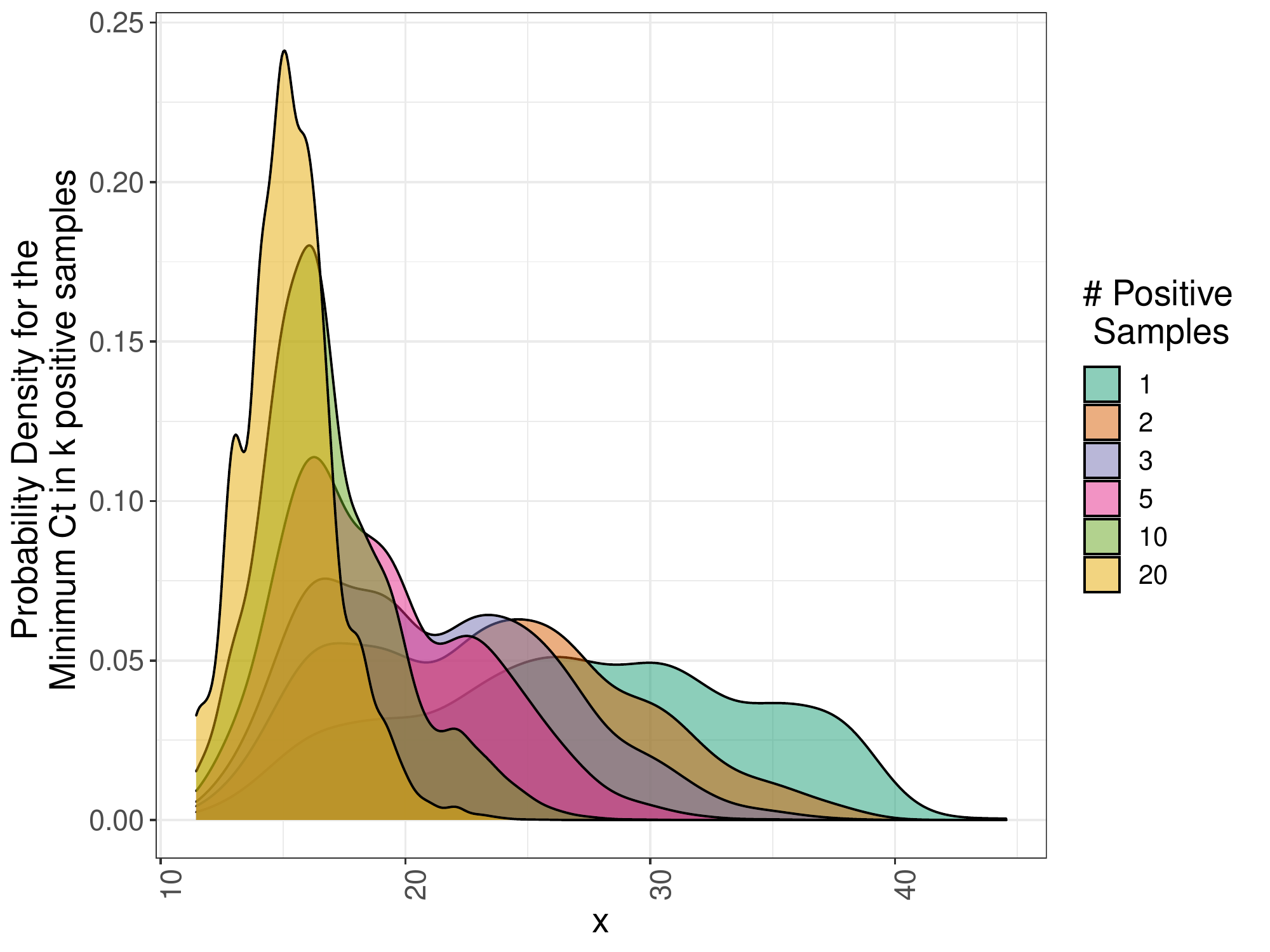}
    \caption{Distribution of the minimum $C_t$}
    \label{fig:min_ct}
     \end{subfigure}  
     \begin{subfigure}[t]{0.5\textwidth}
    \centering
    \includegraphics[width=\textwidth, height=5cm]{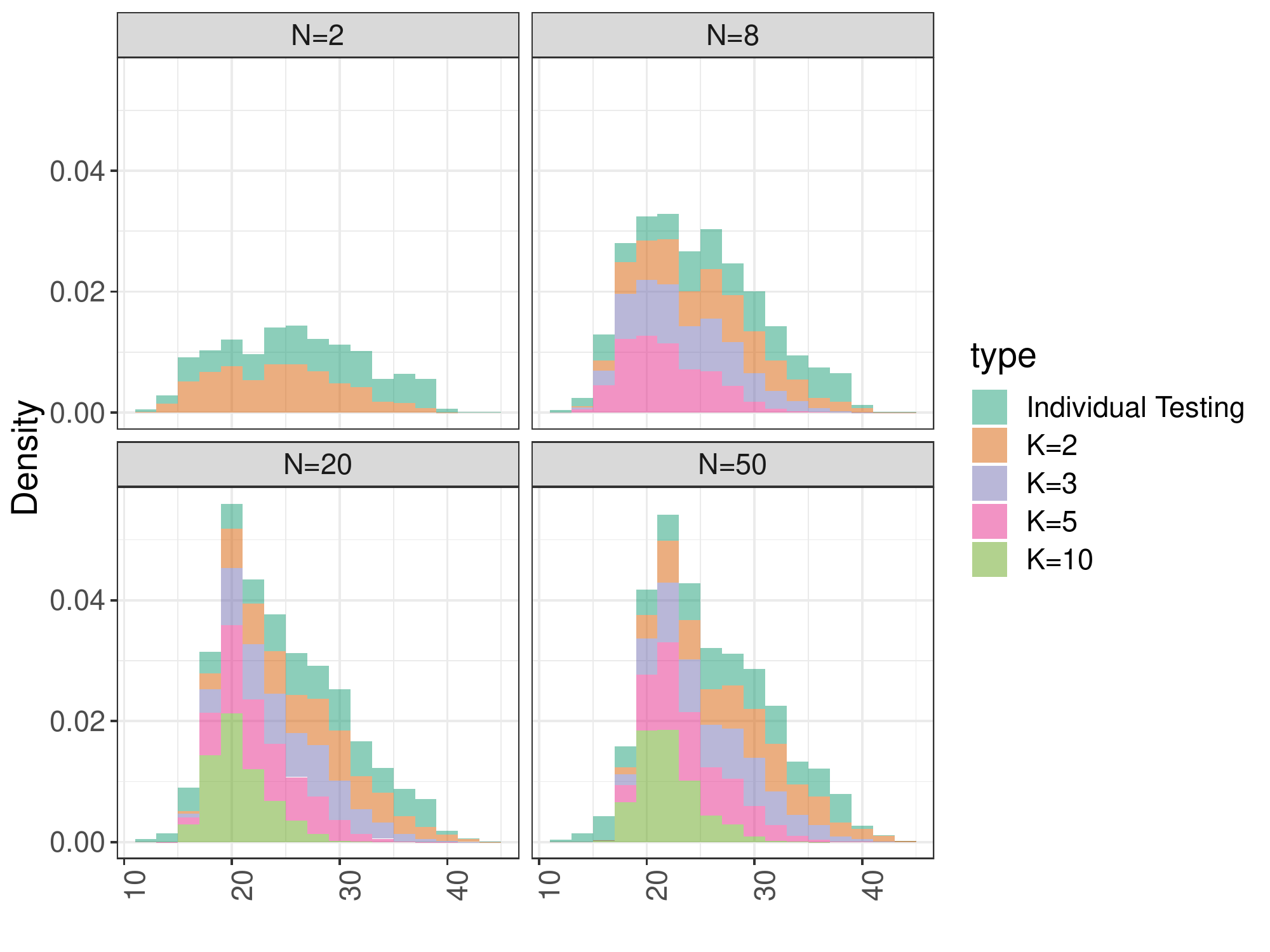}
    \caption{Ct value of the dilution}
    \label{fig:dilution}
 \end{subfigure} 
     \caption{Effect of the dilution on the $C_t$ value}
\end{figure}

Consequently, as shown in Eq. \ref{eq:dilution_ct},  the sensitivity of the pooled test is a function (through the $\text{C}_t$ value) of the number of positive tests per sample. Note that while the samples' infection status $Y$ are correlated, given the $Y_i$'s, the $\text{C}_t$ values can themselves be considered as independent --- that is, there is no evidence (at least, at the time of writing) of the value of the $\text{C}_t$ depending on context (e.g. a sample's corresponding age, gender, genetics, or virus strain). To study the efficiency of this procedure, we consider the  following  two sufficient and mutually exclusive scenarios:
 
 \begin{itemize}
  \setlength\itemsep{1em}
     \item \underline{For $\sum_{i=1}^n Y_i=1$ infected sample in the pool} (which, without loss of generality, we assume to be the first sample), the $\text{C}_t$ of the dilution is distributed as $ C^{\text{dilution}}_t  \overset{D}{=} C^{(1)}_t + \log_2(n)  $: i.e. the dilution translates the distribution of the $\text{C}_t$ value by $\log_2(n)$, and the impact on the sensitivity can be computed directly by translating the sensitivity curve by  $\log_2(n)$. As highlighted by \cite{Barak2021}, for a limit of detection for an individual $\text{C}_t$ value of 40, this means that pooled samples with only one positive sample could only be detected for samples under a $\text{C}_t$ value of $40 - \log_2(n)$; for example, with $n=8$, this would mean a limit of detection corresponding to a $\text{C}_t$ of 37, or for a pool size of $n=20$, a $\text{C}_t$ of 35. It is thus important to assess the stringency of these thresholds in the population being tested. To illustrate this, Figure \ref{fig:lod} provides a visualization of the evolution of the limit of detection and proportion of samples above the limit of detection in a real data set, corresponding to $\text{C}_t$ from positive samples from a Stanford hospital (see Appendix \ref{sec:appendix_sim}). In this case, the proportion of positive samples above the limit of detection $40 - \log_2(n)$ stays below 15\% (thus yielding a sensitivity around the CDC guideline of 85\%, since the sensitivity of the PCR is above 99\% for $C_t <34$) for pool sizes up to size 30. As such, in this particular population, the dilution effect is not overly detrimental to the accuracy of the test.  \cite{Barak2021} further argue that, due to  dwindling infectiousness as $\text{C}_t$ values increase, falsely negative samples with $\text{C}_t$ values 37 and above are less consequential than samples with lower $\text{C}_t$ --- thus mitigating the impact of the dilution on the efficiency of the testing procedure.
     
     \item \underline{For $\sum_{i=1}^n Y_i = k \geq 2$:}  Assuming without loss of generality that the $k$ first samples are positive, we have, by Taylor expansion around the minimum $\text{C}_t$:
     $$C^{\text{dilution}}_t  \overset{D}{\approx} \min_{j \in [1,k]}{C_t^{(j)}} + \log_2(n)  + \frac{1}{\log(2)} \sum_{i\neq \text{argmin}C_t^{(j)} }   2^{ \min_{j \in [1,k]}{C_t^{(j)}} -C_{t}^{(i)}}.$$
     (See Appendix \ref{sec:appendix_comp} for details.) Since the distribution of the $\text{C}_t$ values is heavily skewed, $2^{ \min_{j \in [1,k]}{C_t^{(j)}} -C_{t}^{(i)}} << 1$ with high probability so that $C^{\text{dilution}}_t  \overset{D}{=} \min_{j \in [1,k]}{C_t^{(j)}} + \log_2(n)$  with a good approximation. We provide a visualization of the distribution of the minimum $\text{C}_t$ over a random sample of $k$ $\text{C}_t$ values in our example population in Figure \ref{fig:min_ct}. While this figure uses data from a specific population, we expect this phenomenon to generalize across other populations, since $\text{C}_t$ values are known to exhibit important spread (\cite{Tso2021}). 
     (see Appendix \ref{sec:appendix_comp}). This fact is simply a formalization of the observations by \cite{Barak2021} that the $\text{C}_t$ value of the pooled sample is dominated by the value of the minimum. Thus, for a weakly positive sample (high $\text{C}_t$) to be detected, it is sufficient for it to be combined with a strongly positive sample. We further add to this argument that as the number of positive samples $k$ in the pool increases,  we expect the distribution of the dilution to be increasingly small (and eventually counterbalance the  $\log_2(n)$ offset).  
     To quantify the hitchhiker effect under correlation, we use Monte Carlo simulations to model the behavior of the dilution's $\text{C}_t$. We display the results in Figure \ref{fig:dilution}, where we show the distribution of the $\text{C}_t$ values in our reference population in various dilution effects, and different number of positive samples per pool. Note that for as few as $k=2$ samples, the distribution of the $\text{C}_t$ is comparable to the distribution of individual $\text{C}_t$, and, for these pool sizes, having three positive samples in the pool yields a $\text{C}_t$ distribution for the dilution with a smaller mode. 
 \end{itemize}
 
 The added benefit of this simple formalization is to allow us to study the hitchhiker effect when the independence assumption is violated. Let us consider the low prevalence setting of Assumption 1 ($\pi n <0.10$), and summarize these results as follows.\\
 
 \noindent \textbf{Observation 4.}\\
\textbf{ \underline{(a) i.i.d Case}} For a pool size $n$, by the total law of probability formula, the $\text{C}_t$ value of the dilution is a mixture of n distributions: 
       $$C_t \overset{D}{\approx} -\log_2(n) +  \sum_{k=1}^n 1_{\sum_{i=1}^n Y_i = k}  \min_{1\leq j \leq k} {C_t^{(i)}}$$
       Since in this setting, given that the pool is positive, the probability of having more than two samples is below 0.01, we have:
       $$C_t \overset{D}{\approx} -\log_2(n) +  C_t^{(1)}$$ so that the $\text{C}_t$ value behaves like a translated curve.\\
      \textbf{ \underline{(b) Network effects}} Assuming positive correlations between samples, by Observation 3, we know that we expect $(n-1)\tau$ more samples than under the null hypothesis. This means that the distribution of the minimum $\text{C}_t$ is more heavily skewed towards strongly positive samples, thus ensuring a better probability of detection in the pool.  In fact, we show in Appendix \ref{sec:appendix_comp} Eq.\ref{eq:ct} that the probability that a pooled $\text{C}_t$ with $K$ positive samples exceeds the average individual $\text{C}_t$ value is given by:
      $$  \mathbb{P} [ \min_{i \leq K} (C_{t}^{(i)}) + \log_2(n) > \mathbb{E}[C_t^{(i)} ]   ]  \leq  0.52^{K}  (1 + 0.11 K \log_2(n))  $$
 Thus the probability that the $\text{C}_t$ value of the mixture is greater than the expected $\text{C}_t$ of the individual samples is less than 0.67 for $K=2$, and 0.45 for $K=3$ (assuming worse case $n=100$). Thus, after $K=2$ to $3$ positive samples, the hitchhiker effect ensures that the $\text{C}_t$ value of the mixture will be favorable. While this result uses data from our reference study, we expect the behaviour exhibited here to generalize across populations.

To summarize, in this section, we have shown through simple derivations, backed by simulations, that the dilution only induces  a significant drop in sensitivity if a single sample is positive, but that this effect shrinks as the number of positive samples in the pool increases --- thus confirming and quantifying empirical observations made by \cite{Barak2021}. This allows us to conclude:\\
\xhdr{Lesson 2} \textit{Simple models tailored to the relevant statistical assumptions are useful. The realization that correlated/clustered specimen sampling could yield such considerable gains in $\text{C}_t$ value is only possible through a careful consideration of practical field constraints and observed measurements.}
     
\section{Evaluating Practical Efficiency}\label{sec:metrics}

\subsection{Metrics of Interest and Benchmarks} 

Having studied the impact of the network transmission on the amount of viral material present in the sample, we now assess the impact of these departures from the traditional framework on our metrics of interest. To measure the efficiency of pooled testing under correlations, we introduce a set of four performance metrics, consistent with the group testing literature: the sensitivity ($s$), relative sensitivity ($s_r$) (which compares individual to pooled testing procedures), expected number of tests per sample ($\eta$), and proportion of missed cases per sample. 

\xhdr{Sensitivity ($s$) and Relative Sensitivity ($s_r$)} Sensitivity ($s$) is the proportion of true-positives that are detected by the test, that is, the probability of a positive test result given that there is at least one positive sample in the pool. 

\begin{equation*}
    \begin{split}
        \textbf{Sensitivity: } s &= \mathbb{P}[\text{test is positive} | \sum_{i=1}^n Y_i \ge 1]\\
        &= \frac{\sum_{k=1}^n \mathbb{P}[\text{test is positive} \bigcap \{ \sum_{i=1}^n Y_i=k\}] }{\mathbb{P}[\sum_{i=1}^n Y_i \ge 1]} \\
        &= \frac{\sum_{k=1}^n \mathbb{P}[\text{test is positive} | \sum_{i=1}^n Y_i=k] p_k}{\mathbb{P}[\sum_{i=1}^n Y_i \ge 1]},
    \end{split}
\end{equation*}
with $p_k = \mathbb{P}[\sum_{i=1}^n Y_i = k]$. We assume that specificity (probability of a false positive) is zero.

However, tests are inherently imperfect, and such sensitivity might not be realistically achievable. As such, a more informative metric is the sensitivity of the pooled procedure, compared to individual testing (which we consider to be our ``gold-standard''). Relative sensitivity serves as a comparison of the probability of a positive test result in a pooled testing scenario to an individual testing scenario which might differ as a product of the dilution effect. 
\begin{equation*}
    \begin{split}
        \textbf{Relative Sensitivity: } s_r 
        &= \frac{\mathbb{P}[\text{test is positive} | \sum_{i=1}^n Y_i \ge 1]}{\mathbb{P}[\text{Individual test is positive} |Y_1  = 1]}\\
    \end{split}
\end{equation*}
Contrary to the sensitivity, this measure can take values in $\mathbb{R}^+:$ a relative sensitivity lower than one indicates a lower sensitivity of the pooled procedure relative to individual testing, whereas a value of $s_r$ greater than one would indicate a better sensitivity in the case of pooled testing with respect to individual testing.

Both sensitivities can be decomposed as a function of number of positive samples per pool and $\text{C}_t$ value as follows:
\begin{equation*}
    \begin{split}
    \forall x, \quad  \mathbb{P}[s > x]  &=  \int \mathbb{P}[s > x | C_t] p(C_t) dCt =   \int \sum_{k=1}^{n} \mathbb{P}[s > x | C_t] p(C_t, \sum_{i=1}^n Y_i = k) dCt \\
  &=   \sum_{k=1}^{n}  \mathbb{P}[ \sum_{i=1}^n Y_i = k]  \underbrace{\int \mathbb{P}[s > x | C_t] p(C_t|  \sum_{i=1}^n Y_i = k) dCt}_{=s_k, \text{ by definition}} \\
   &=   \sum_{k=1}^{n}  p_k   s_k \\
    \end{split}
\end{equation*}
where we have considered here the sensitivity of the test $s_k$ to be a function of the number of positives (marginalized over $C_t$ values). In light of our discussion of the effect of the network effects on the $\text{C}_t$ value of the dilution, we conclude that the existence of correlations has a positive effect on the sensitivities.

\xhdr{Expected Number of Tests per Sample ($\eta$)} This measures the efficiency of the pooled testing procedure. Since in the two-step Dorfman procedure every sample has to be re-tested if the pool is tested positive, the efficiency is:
\begin{equation*}
    \begin{split}
    \eta &=  \frac{1}{n} \times \mathbb{E}[1 + n \times \mathbb{P}[\text{test is positive}]  ] = \frac{1}{n} + \mathbb{P}[\text{test is positive}]\\
    &= \frac{1}{n} + \sum_{k=1}^n \mathbb{P}[\text{test is positive} |\sum_i Y_i =k] p_k
    \end{split}
\end{equation*}
This measure has to be compared against the benchmark value of $\eta_0 = 1$, which is the efficiency of the individual testing procedure (pool size $n=1$). However, it is important to note that the expected number of tests per sample must be considered in conjunction with other metrics, as it is only a partial indicator of the validity of the procedure. Indeed, a faulty test which is always negative will achieve the best efficacy $\eta=\frac{1}{n}$, but with zero sensitivity. 

 \xhdr{Proportion of Missed Cases per Sample} We also consider the proportion of cases that the grouped testing procedure fails to detect per test.
 $$=   \frac{\sum_{k=1}^n  k \mathbb{P}[\text{test is positive} | \sum_{i=1}^n Y_i=k] (1-p_k)}{\frac{1}{n} + \mathbb{P}[\text{test is positive}]}$$

\xhdr{Simulations} To illustrate our analysis, we perform Monte Carlo simulations and calculate our metrics of interest at varying pool sizes and values of $\pi$ and $\tau$.  We simulate two-stage pooled testing setting where individuals' infection statuses (positive vs. negative) are correlated, and compare this to our null model of assuming uncorrelated individuals. To make our analysis more realistic, the values of $\pi$ and $\tau$ are informed by fitting Beta distributions to published literature and data. For $\tau$, we fit Beta distributions to SAR values reported for a range of settings, including households with symptomatic index cases, households with asymptomatic index cases, transmission between spouses, healthcare settings, and from a child index case. For $\pi$, we fit Beta distributions to prevalence data at differing times over the course of the pandemic and geographic locations in the United States. The chosen time points and geographic locations are intended to be representative of varying prevalence levels and stages of the pandemic (e.g. rising cases, falling cases, etc). Details of the methodology for fitting the distributions, as well as information on the settings and distribution parameters are described in Appendix \ref{sec:appendix_sim}. Pool sizes range from $n$ of 1 to $n$ of 30.\\
\indent  To compute the sensitivity of the PCR test given a pool of size $n$ containing $1,...,n$ positive samples, we use empirically collected data on the distribution of $\text{C}_t$ values from \cite{Wang2021}, which is best represented by a Weibull distribution with shape parameter $s=4.5$ and scale $\eta = 30$. The distribution of the $\text{C}_t$ values depends on a number of factors, including the population tested (ie., hospital admissions vs general population, COVID variant, etc). To create a realistic distribution of $\text{C}_t$ values with the appropriate amount of spread, we sample and shift the Weibull distribution of \citeauthor{Wang2021}: we sample from their fitted distribution to create a mock distribution of individual $\text{C}_t$ values, and shift it to model a population in which 30\% of samples are above $\text{C}_t= 35$. \\
\indent For each combination of $\pi$, $\tau$, and pool size, we calculate the metrics of interest (sensitivity, relative sensitivity, expected number of tests per sample, missed cases per sample), weighted by the probability of observing $k$ positives in that particular pool. We simulate the situation of testing a population of correlated individuals, where we either ignore correlation and erroneously treat the individuals as independent ($H_0$), or correctly consider networks of correlated individuals ($H_a$). In the null model, we make i.i.d assumptions and the expected number of positives per pool has a binomial distribution (probability of observing $K$ positives in $n$ trials (pool size) with success probability $\pi$ and $\tau = 0$. For the alternative model $H_a$, the probability of $K$ positives is computed exactly using the Poisson Binomial distribution (Appendix \ref{sec:appendix_sim}). 

 \begin{figure}
    \includegraphics[width=\linewidth]{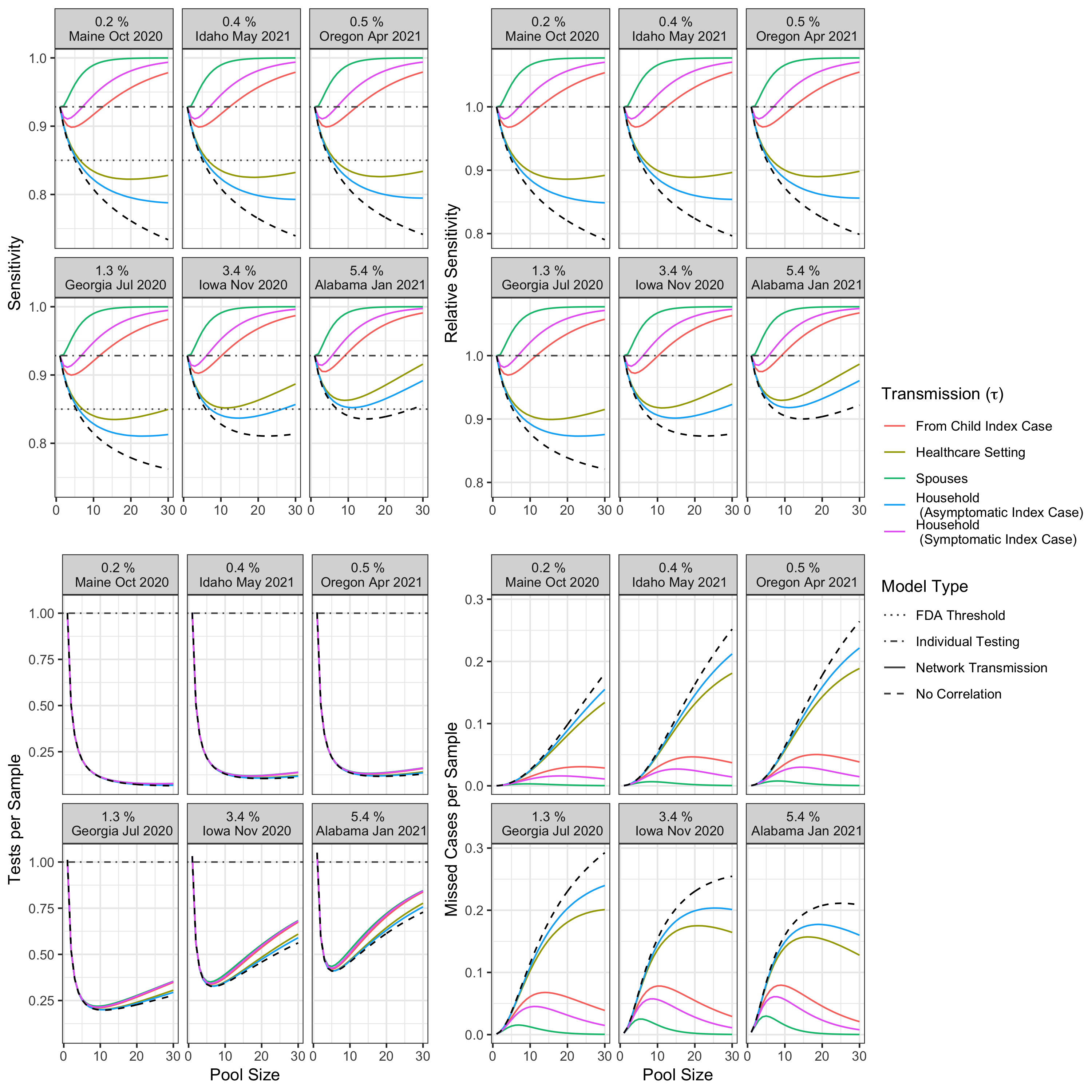}\hfill
  \caption{\textbf{Fixed Model}: Model-estimated parameters (sensitivity, relative sensitivity, expected tests per sample, missed cases per sample) by pool size, prevalence ($\pi$), and network transmission probability ($\tau$). Null (no correlation) model, individual testing, and FDA sensitivity threshold are also indicated, where relevant.}
   \label{fig:allmetrics_fixed}
\end{figure}

\xhdr{Discussion of Simulation Results}  Across all prevalence settings and levels of network transmission, the model that accounts for network transmission (correlations in pools) performs better in terms of higher sensitivity and fewer missed cases per sample than the null model (which ignores correlations) (Figure ~\ref{fig:allmetrics_fixed}). Accounting for correlations between individuals can result in large percentage increases in sensitivity over the null model; for example, for spousal network transmission in both low prevalence (e.g. Maine October 2020) and high prevalence (e.g. Alabama January 2021) settings, we observe 31.25\% and 19.14\% increases in sensitivity, respectively, compared to the null model (Table~\ref{table:summary_performance_fixed}). 

Comparing the sensitivity of the pooled procedure to individual testing, high levels of network transmission (such as observed between spouses and in a household with a symptomatic index case) results in sensitivity greater than the individual test, and far exceeding the minimum FDA threshold (0.85). In low prevalence settings or weak network transmission (such as healthcare settings or households with an asymptomatic index case), the sensitivity of the pooled testing procedure may fall below the FDA threshold (0.85) at large pool sizes. At sufficiently high prevalence levels (such as observed in Alabama in January 2021), pools of all sizes (including as large as 30) exceed the FDA threshold for pooled testing sensitivity. 

The pooled procedure also results in large decreases in the number of tests needed per sample, when compared to individual testing. Implementing pooled testing in Maine during October 2020 (low prevalence) among households with asymptomatic index cases, households with symptomatic index cases, and spouses could reduce the number of tests per sample by over 92\% in all three network transmission settings (Table ~\ref{table:summary_performance_fixed}). In higher prevalence settings, reductions in testing associated with pooled testing are more modest, but still upwards of 20\% (Table ~\ref{table:summary_performance_fixed}). 

From these results, we draw the following conclusion:\\
\xhdr{Lesson 3} \textit{The utility of pooled testing is context dependent, but statistical models informed by observed data in a range of prevalence and network settings demonstrate that accounting for non-i.i.d settings uniformly improves the expected performance of the procedure.}

\begin{table}
\caption{Expected performance of pooled testing for the fixed model under select prevalence, network transmission, and pool size scenarios.}
\label{table:summary_performance_fixed}
\setlength{\tabcolsep}{10pt} 
\renewcommand{\arraystretch}{1.1}
\resizebox{\textwidth}{!}{\begin{tabular}{@{}ccccc@{}} 
\hline
\multicolumn{3}{p{3.5cm}}{ }&
\multicolumn{2}{p{14cm}}{\centering \textbf{Fixed Model}}\\
 \cline{4-5}
 &   Network Transmission ($\tau$) & Prevalence ($\pi$)& 
 \multicolumn{1}{p{7cm}}{\centering \% Increase Sensitivity \\ Relative to $H_0$}&
\multicolumn{1}{p{7cm}}{\centering \% Decrease Tests/Sample \\ Relative to Individual Testing}\\
 \hline
\multirow{6}{*}{\rotatebox[origin=c]{90}{Pool Size n = 5}} 
& \multirow{2}{*}{Household Asympt. (0.012)}  
& Maine Oct. 2020 (0.2\%) &  0.62 &  79.35\\
& & Alabama Jan. 2021 (5.4\%) & 0.57 &  58.92\\[10pt]

& \multirow{2}{*}{Household Symptomatic (0.18)}  
&  Maine Oct. 2020 (0.2\%)  & 7.77 &  79.31\\
& &  Alabama Jan. 2021 (5.4\%) & 7.11 & 57.54\\[10pt]

& \multirow{2}{*}{Spouses (0.38)} 
&  Maine Oct. 2020 (0.2\%) & 13.02 & 79.28\\
& &  Alabama Jan. 2021 (5.4\%) & 11.85 & 56.55\\
 \hline
 
 \multirow{6}{*}{\rotatebox[origin=c]{90}{Pool Size n = 20}}
& \multirow{2}{*}{Household Asympt. (0.012)}  
& Maine Oct. 2020 (0.2\%) & 4.46 &  92.63\\
& & Alabama Jan. 2021 (5.4\%) & 3.12 &  36.88\\[10pt]
 
& \multirow{2}{*}{Household Symptomatic (0.18)}  
&  Maine Oct. 2020 (0.2\%)  & 28.85 &  92.07\\
& &  Alabama Jan. 2021 (5.4\%) & 17.87 & 28.57 \\[10pt]

& \multirow{2}{*}{Spouses (0.38)} 
&  Maine Oct. 2020 (0.2\%) & 31.25 &  92.02 \\
& &  Alabama Jan. 2021 (5.4\%) & 19.14 &  27.86 \\
 \hline
\end{tabular}}
\footnotesize{Percent increase in sensitivity relative to $H_0$ = $100 *(Sens_{H_a} - Sens_{H_0})/(Sens_{H_0})$; \\ Percent decrease in tests per sample relative to individual testing = $100 * (Tests_{H_a} - n)/(n)$, ($n$=pool size).}
\end{table}

\section{Testing under Uncertainty} \label{sec:heterogeeity}

So far, our discussion has focused  on situations where the prevalence and the transmission levels are fixed, known quantities.  As such, all the quantities that we have computed are conditional expectations given $\tau, \pi$. However, in practice, these are estimates with associated levels of uncertainty and thus can themselves be modeled  as random variables, whose variability has to be taken into account as we compute metrics of interest. Given the non-linear nature of the model and the wide uncertainty around the value of the network transmission rate (or SAR) $\tau$, it is important to evaluate how much this added variability affects our estimates of the performance of pooled sampling. Uncertainty and heterogeneity are crucial aspects of COVID-19 kinetics that need to be accounted for to ensure accurate  epidemiological predictions  (\cite{cirillo2020tail,cave2020covid,gomez2020mapping,zhang2020evaluating}): most COVID-19 forecasting models --- whether geared towards the prediction of the incidence rate, underascertainment bias, or towards the performance of pooled testing, such as the one considered in this paper ---  are indeed non-linear functions of many unknown and/or highly variable quantities. When solely considering the average rather than accounting for the distributional nature of these variables,  the error can rapidly amplify, and thus needs to be appropriately characterized and controlled (\cite{donnat2021modeling}).  In this section, we (a) show that in the prevalence/SAR regimes that we are considering,  the main driver of the heterogeneity lies in the uncertainty around network transmission and is a function of the behavior of the distribution at the tail, rather than of its variability, (b) highlight ranges of parameters $(n, \pi, \tau)$ which are robust to this heterogeneity, and (c) show via experiments how to construct (and interpret) prediction intervals for the performance of pooled sampling under uncertainty.

\xhdr{Observation 5} \textit{The sensitivity and efficiency (tests per sample) of the alternative model can never be worse than the null model.} This is due to the fact that the (true) effective number of tests per samples can be written as:
$$\eta = \frac{1 +  \mathbb{P}[\sum_{i=1}^n Y_i>0]}{n} =\frac{1}{n} + \frac{1- (1-\pi)^n }{n} =\frac{1}{n} + \pi + o(\pi)$$ in both scenarios. From this formulation we observe that the efficiency is not a function of the network transmission, and depends only on community transmission rates. As such, results on the efficiency are robust for all parameterizations and levels of uncertainty in the value of $\tau$, but will solely depend on the uncertainty for $\pi$. This fact also highlights the necessity of having accurate estimates of the prevalence, tailored to the population at hand in order to correctly optimize pooled testing. In this context, estimates of the prevalence for the sampling population, using hyper-local data and/or additional covariates such as vaccination rates  can be crucial in  further reducing  this uncertainty (\cite{harvard2021,cramer2021evaluation,donnat2021predictive, zhou2020spatiotemporal}).

For the sensitivity, since correlations can only increase the number of positive samples per pool and sensitivity is an increasing function of the $C_t$, the sensitivity can only be improved by taking into account correlations between individuals in the alternative model.

\xhdr{Observation 6} \textit{We can identify settings in which pooled sampling will have worse sensitivity than individual testing.} Since under correlations, the number of positive samples per infected pool is  well approximated by $1 + \text{Binomial}(n-1, \tau)$ (since with 95\% probability, community transmission yields only one infected sample), the number of infected samples can vary quite substantially:
\begin{equation*}
    \begin{split}
    \mathbb{V}_{H_a}[\sum_{i=1}^n Y_i | \sum_{i=1}^n Y_i>0]& = \mathbb{E}[ \mathbb{V}_{H_a}[\sum_{i=1}^n Y_i | \tau, \sum_{i=1}^n Y_i>0]] + \mathbb{V}[ \mathbb{E}_{H_a}[\sum_{i=1}^n Y_i | \tau, \sum_{i=1}^n Y_i>0]]    \\
    & \approx \mathbb{E}[(n-1)\tau(1-\tau)] + \mathbb{V}[ 1 + \tau (n-1)]    \\
      & \approx  (n-1) (n-2) \sigma_{\tau}^2   + (n-1)\mu_{\tau}(1-\mu_{\tau})  \\
    \end{split}
\end{equation*}
where $\mu_{\tau}$ and $\sigma_{\tau^2}$ are respectively the mean and variance of $\tau$.  We can now examine the effect of this variance on the sensitivity.
As seen in the first section, the sensitivity is impacted if the number of positive samples in infected pools falls below 2. This would mean that network transmission only accounts for a single additional infected sample in the pool. 
This event, which has (by property of the binomial) probability equal to  $(n-1) \tau(1-\tau)^{n-2}$ can be deemed highly unlikely as long as it happens with probability less than 0.05. This allows us to solve for regions $\Omega$ of the parameter space for $\tau$ where we expect robust performances of pooled testing, which improve upon individual testing:
\begin{equation*}
    \begin{split}
    \Omega = \{ \tau \in[0,1]: \quad   (n-1) \tau(1-\tau)^{n-2} \leq 0.05 \}.
        \end{split}
\end{equation*}
This allows the user to select a pool size such that, with probability $ 1- \mathbb{P}[\tau \in \Omega]$, the procedure is better than individual testing. Choosing the value of $\eta$ is setting a tolerance threshold, and should be defined by the user.



\xhdr{Experiments}

We adapt our simulations in the previous section to account for heterogeneity and uncertainty in $\pi$ and $\tau$ by performing the simulations under three further settings: 
\begin{description}
    \item[Setting 1 - Fixed $\pi$ / Random $\tau$ ](Fixed Prevalence/Random Network Effect): Sample $\tau$ from a Beta prior distribution  and calculate the corresponding probability of $k$ positives due to network transmission. $\pi$ assumed fixed and equal to the point estimate. 
  \item[Setting 2 - Random $\pi$ / Fixed $\tau$] (Random Prevalence/Fixed Network Effect).   Sample $\pi$ from a Beta prior distribution. $\tau$ assumed fixed and equal to the point estimate.
\item[Setting 3 - Random $\pi$ / Random $\tau$ ](Random Prevalence/Random Network Effect): Sample both $\pi$ and $\tau$, as described above.
\end{description}
When sampling $\pi$ and $\tau$, for each point estimate and group size, we perform $B=100$ simulations of sampling from the prior distribution. 
 
\begin{figure}[H]
    \includegraphics[width=\linewidth]{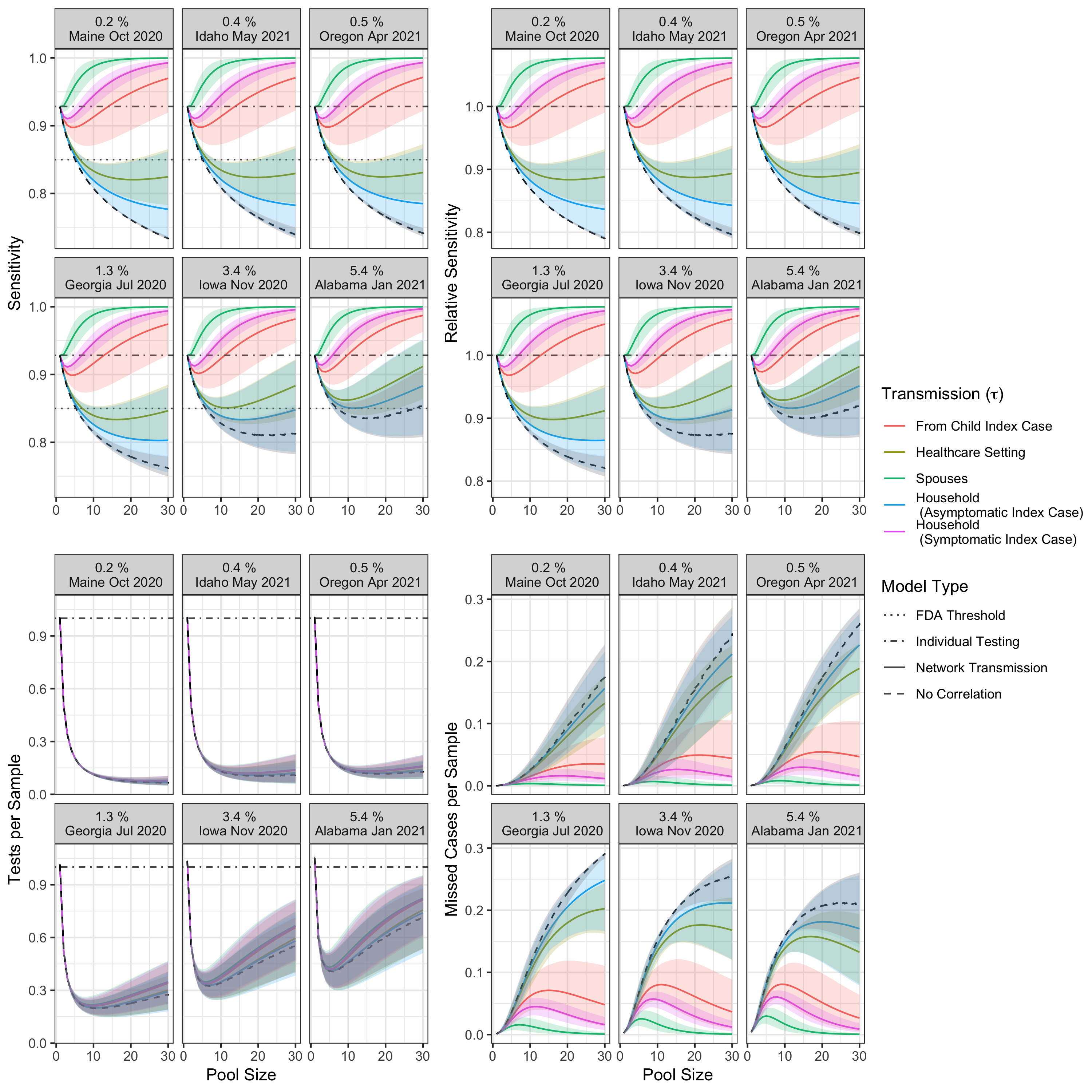}\hfill
  \caption{\textbf{Random Prevalence/Random Network Effect Model}: Model-estimated parameters (sensitivity, relative sensitivity, expected tests per sample, missed cases per sample) by pool size, prevalence ($\pi$), and network transmission ($\tau$) when $\pi$ and $\tau$ are sampled from Beta distributions corresponding to the specified scenario. Shaded area indicates the empirical 95 percent credible prediction interval. Proportion of samples with $C_t$ above the 95\% LoD (held constant at LoD = 35) is constant, equal to 25\%. The null (no correlation) model, individual testing, and FDA sensitivity threshold are also indicated, where relevant.}
   \label{fig:allmetrics_allgraph}
\end{figure}

 \indent The results of the Monte Carlo simulations for Random $\tau$ and $\pi$ (Random Prevalence/Random Network Effect) are shown in Figure \ref{fig:allmetrics_allgraph}; results for random $\tau$ (Fixed Prevalence/Random Network Effect) and random $\pi$ (Random Prevalence/Fixed Network Effect) are presented in Appendix ~\ref{sec:appendix_sim}, Figures ~\ref{fig:allmetrics_taugraph} and ~\ref{fig:allmetrics_pigraph}. 
 
 These simulations inform the following observation: \\
 \xhdr{Lesson 4} \textit{These findings highlight the importance of developing \textbf{adaptive} pooled testing procedures. With reasonably specified models for prevalence and network transmission, the optimal pool size can be chosen to maximize sensitivity and ensure minimum FDA thresholds are met.  However, even with significant uncertainty and heterogeneity in pools, results are robust. }
 
 \begin{figure}
\includegraphics[width=\linewidth]{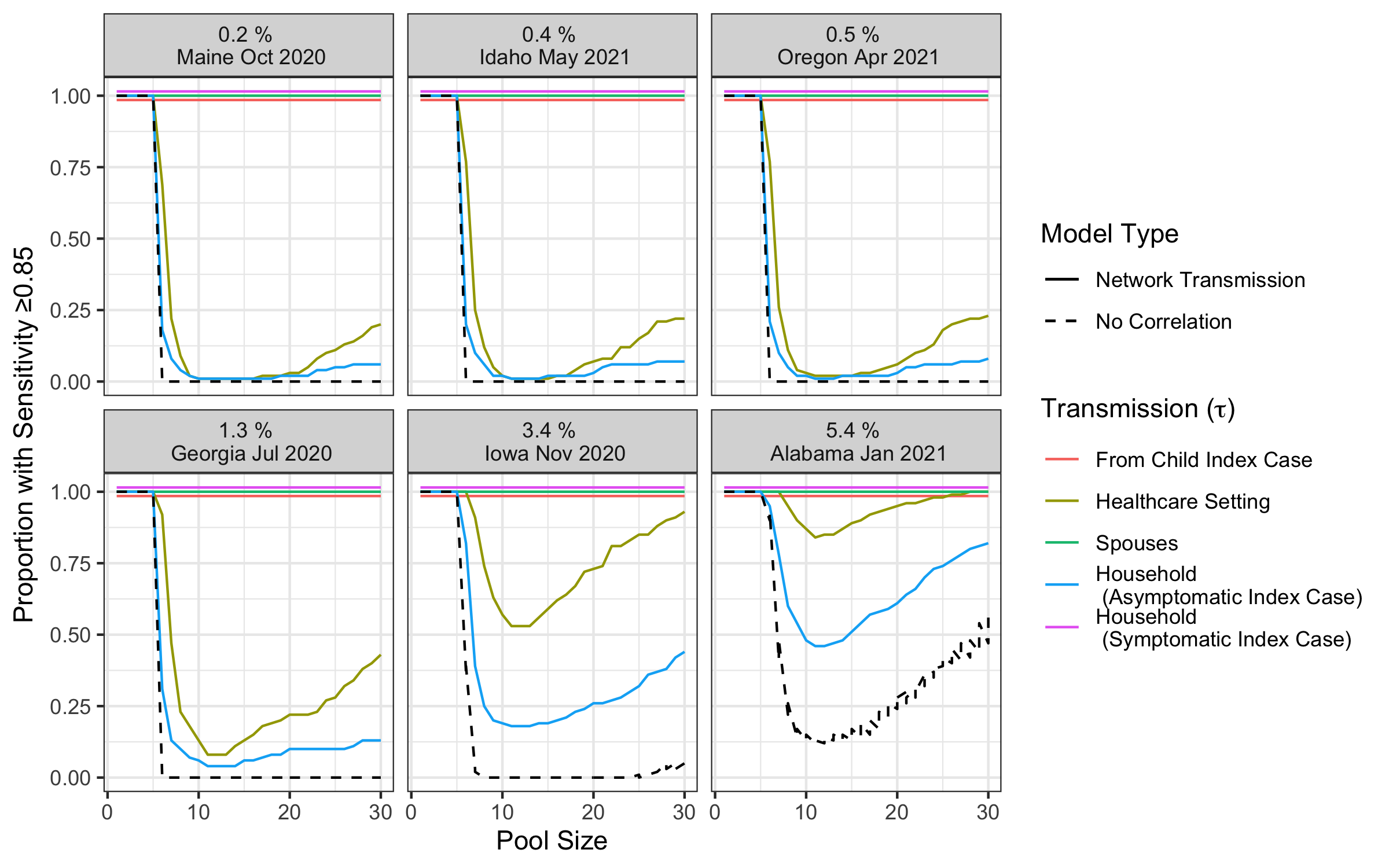}\hfill
\caption{Proportion of simulated results with sensitivity equal to or greater than the FDA threshold for pooled testing (0.85), using the Random Prevalence/Random Network Effect Model.}
\label{fig:sens_fda}
\end{figure}
 
 In settings with high network transmission (e.g. child index cases, spouses, and household symptomatic index cases), all of the simulation results have sensitivity greater than the FDA threshold for sensitivity (Figure ~\ref{fig:sens_fda}). For low network transmission settings (e.g. healthcare or household asymptomatic index cases), pooled testing with small pools ($n \leq 5$) may still meet minimum FDA standards, but larger pools may not be appropriate (Figure ~\ref{fig:sens_fda}). 
 Supporting our previous observations, heterogeneity in results is primarily driven by uncertainty in $\tau$, not in $\pi$. 
 
Additionally, adaptability of the overall model is critical in a pandemic setting, where transmissibility and susceptibility vary over time and space as a function of the particular viral variants circulating and the prevalence of vaccination (and efficacy of vaccines against variants). The simulations presented in this paper can easily be adapted to settings where more transmissible variants are widespread (by increasing the value of the transmission parameter) or as vaccination rates increase (by decreasing the value of the prevalence parameter). The robustness of results to significant uncertainty in parameter values is especially critical in this setting.

 Considering the efficiency of the pooled testing procedure, there is very little variation in expected tests per sample across different specifications of the model and parameters (Figures \ref{fig:allmetrics_allgraph}; Table \ref{table:summary_performance}). In all cases, the pooled sampling procedure results in significant reductions in tests per sample relative to the individual sampling procedure.

\begin{table}
\caption{Expected performance of pooled testing under select prevalence, network transmission, model type, and pool size scenarios.} 
\label{table:summary_performance}
\setlength{\tabcolsep}{10pt} 
\renewcommand{\arraystretch}{1.1}
\resizebox{\textwidth}{!}{\begin{tabular}{@{}ccccccc@{}} 
\hline
\multicolumn{3}{p{3.5cm}}{ }&
\multicolumn{4}{p{8cm}}{\centering \textbf{Model}}\\
 \cline{4-7}
 \multicolumn{3}{p{2.5cm}}{ }&
\multicolumn{2}{p{4cm}}{\centering Random Prevalence/ Fixed Network Transmission}&
\multicolumn{2}{p{4cm}}{\centering Random Prevalence/ Random Network Transmission}\\
 \cline{4-7}
 & Network Transmission ($\tau$)  &Prevalence ($\pi$) & 
 \multicolumn{1}{p{2cm}}{\centering \% Increase Sensitivity \\ Relative to $H_0$}&
\multicolumn{1}{p{2cm}}{\centering \% Decrease Tests/Sample \\ Relative to Individual Testing}&
\multicolumn{1}{p{2cm}}{\centering \% Increase Sensitivity \\ Relative to $H_0$}&
\multicolumn{1}{p{2cm}}{\centering \% Decrease Tests/Sample \\ Relative to Individual Testing}\\
 \hline
\multirow{6}{*}{\rotatebox[origin=c]{90}{Pool Size n = 5}} 
& \multirow{2}{*}{Household Asympt. (0.012)}  
& Maine Oct. 2020 (0.2\%) &  0.52 &  79.34 & 0.58 & 79.36\\
& & Alabama Jan. 2021 (5.4\%) & 0.50 &  58.77 & 0.53 &  58.92\\[10pt]
& \multirow{2}{*}{Household Symptomatic (0.18)}  
&  Maine Oct. 2020 (0.2\%)  & 7.71 &  79.31 & 7.86 &  79.31\\
& &  Alabama Jan. 2021 (5.4\%) & 7.04 & 58.00 & 7.19 &  57.52\\[10pt]
& \multirow{2}{*}{Spouses (0.38)} 
&  Maine Oct. 2020 (0.2\%) & 12.8 & 79.28 & 13.09 &  79.27\\
& &  Alabama Jan. 2021 (5.4\%) &11.67 & 56.94  & 11.92 &  56.54\\
 \hline
 
 \multirow{6}{*}{\rotatebox[origin=c]{90}{Pool Size n = 20}}
& \multirow{2}{*}{Household Asympt. (0.012)}  
& Maine Oct. 2020 (0.2\%) & 3.62 &  92.60 & 3.94 &  92.64\\
& & Alabama Jan. 2021 (5.4\%) & 2.55 &  37.74 & 2.71 &  37.12\\[10pt]
& \multirow{2}{*}{Household Symptomatic (0.18)}  
&  Maine Oct. 2020 (0.2\%)  & 28.69 &  92.07 & 28.86 & 92.07\\
& &  Alabama Jan. 2021 (5.4\%) &17.76 &   30.51 & 17.88 &  28.57\\[10pt]
& \multirow{2}{*}{Spouses (0.38)} 
&  Maine Oct. 2020 (0.2\%) & 31.17 &  92.03  & 31.20 &  92.02\\
& &  Alabama Jan. 2021 (5.4\%) & 19.09 &  29.70  & 19.11 &  27.87\\
 \hline
\end{tabular}}
\footnotesize{Percent increase in sensitivity relative to $H_0$ = $100 \times (\text{Sens}_{H_a} - \text{Sens}_{H_0})/(\text{Sens}_{H_0})$; \\ Percent decrease in tests per sample relative to individual testing = $100 \times (\text{Tests}_{H_a} - n)/(n)$, where $n$=pool size.}
\end{table}

\section{Conclusion}\label{sec:conclusion}

Violations of i.i.d assumptions and high prevalence settings have typically been considered impediments to using pooled testing; however, we demonstrate, mathematically and via simulation, that significant gains can be made in terms of testing efficiency and sensitivity by taking advantage of correlated samples and high probabilities of positive samples. Clustering multiple positive samples in a single pool improves both efficiency (by focusing re-testing in a directed fashion) and sensitivity (by increasing the concentration of viral genetic material). We also investigate the effects of heterogeneity in the input parameters- specifically prevalence and network transmission- on the outcome metrics of interest. We find that our overall findings are robust to heterogeneity. This suggests that these methods can be useful even in real-world settings where precise values of prevalence and network transmission may not be known.\\
\indent For example, at the peak of the pandemic in Alabama in January 2021, pooled samples of size 5 collected from households with symptomatic index cases could have reduced testing needs by 58\%; at that time, the positive test rate in Alabama exceeded 20\%, and greater testing capacity was urgently needed. Pooling in this setting and with this pool size is predicted to have sensitivity far exceeding individual tests, even when accounting for uncertainty in both prevalence and network transmission rates. Even greater savings can be observed in other settings, such as Maine in October 2020, a period when prevalence was low: pooled testing in pools of size 20 would have reduced testing requirements by over 90\% relative to individual testing in both high and low network transmission settings. For moderately high values of network transmission (household transmission with a symptomatic index case and spouses), the sensitivity of this procedure again exceeds individual testing.\\
In conclusion, as a lesson learned from the pandemic, we highlight the importance of "field-aware" statistical modeling and the importance of adaptive models. To develop an actionable response to the pandemic and its unprecedented conditions, it is important to develop statistical model that accurately optimize procedures to the population at hand. This development has to be done in close collaboration with clinicians, to ensure the feasibility and scalability of the proposed solution. Here, we demonstrate that leveraging correlations in specimen collection procedures and incorporating knowledge about local prevalence and network transmission parameters can lead to better informed, logistically-feasible, and adaptive pooled testing.


\section{Acknowledgements}\label{sec:acknowledgements}
We thank Dr. Benjamin Pinsky (Associate Professor of Pathology and Medicine at Stanford University) for sharing data with us and for his feedback on the manuscript. 

\medskip
\pagebreak

\bibliographystyle{imsart-nameyear} 


\begin{thebibliography}{49}

\bibitem[\protect\citeauthoryear{Abdalhamid
  et~al.}{2020}]{Abdalhamid2020AssessmentResources}
\begin{barticle}[author]
\bauthor{\bsnm{Abdalhamid},~\bfnm{Baha}\binits{B.}},
  \bauthor{\bsnm{Bilder},~\bfnm{Christopher~R}\binits{C.~R.}},
  \bauthor{\bsnm{McCutchen},~\bfnm{Emily~L}\binits{E.~L.}},
  \bauthor{\bsnm{Hinrichs},~\bfnm{Steven~H}\binits{S.~H.}},
  \bauthor{\bsnm{Koepsell},~\bfnm{Scott~A}\binits{S.~A.}} \AND
  \bauthor{\bsnm{Iwen},~\bfnm{Peter~C}\binits{P.~C.}}
(\byear{2020}).
\btitle{{Assessment of Specimen Pooling to Conserve SARS CoV-2 Testing
  Resources}}.
\bjournal{American Journal of Clinical Pathology}
\bvolume{153}
\bpages{715--718}.
\bdoi{10.1093/ajcp/aqaa064}
\end{barticle}
\endbibitem

\bibitem[\protect\citeauthoryear{Adam et~al.}{2020}]{Adam2020ClusteringKong}
\begin{barticle}[author]
\bauthor{\bsnm{Adam},~\bfnm{Dillon~C.}\binits{D.~C.}},
  \bauthor{\bsnm{Wu},~\bfnm{Peng}\binits{P.}},
  \bauthor{\bsnm{Wong},~\bfnm{Jessica~Y.}\binits{J.~Y.}},
  \bauthor{\bsnm{Lau},~\bfnm{Eric H.~Y.}\binits{E.~H.~Y.}},
  \bauthor{\bsnm{Tsang},~\bfnm{Tim~K.}\binits{T.~K.}},
  \bauthor{\bsnm{Cauchemez},~\bfnm{Simon}\binits{S.}},
  \bauthor{\bsnm{Leung},~\bfnm{Gabriel~M.}\binits{G.~M.}} \AND
  \bauthor{\bsnm{Cowling},~\bfnm{Benjamin~J.}\binits{B.~J.}}
(\byear{2020}).
\btitle{{Clustering and superspreading potential of SARS-CoV-2 infections in
  Hong Kong}}.
\bjournal{Nature Medicine}
\bvolume{26}
\bpages{1714--1719}.
\bdoi{10.1038/s41591-020-1092-0}
\end{barticle}
\endbibitem

\bibitem[\protect\citeauthoryear{ASM}{2021}]{asm_testing}
\begin{bmisc}[author]
\bauthor{\bsnm{ASM}}
(\byear{2021}).
\btitle{Supply Shortages Impacting COVID-19 and Non-COVID Testing}.
\bnote{[Online; posted 19-January-2021]}.
\end{bmisc}
\endbibitem

\bibitem[\protect\citeauthoryear{Barak et~al.}{2021}]{Barak2021}
\begin{barticle}[author]
\bauthor{\bsnm{Barak},~\bfnm{Netta}\binits{N.}},
  \bauthor{\bsnm{Ami},~\bfnm{Roni~Ben}\binits{R.~B.}},
  \bauthor{\bsnm{Sido},~\bfnm{Tal}\binits{T.}},
  \bauthor{\bsnm{Perri},~\bfnm{Amir}\binits{A.}},
  \bauthor{\bsnm{Shtoyer},~\bfnm{Aviad}\binits{A.}},
  \bauthor{\bsnm{Rivkin},~\bfnm{Mila}\binits{M.}},
  \bauthor{\bsnm{Licht},~\bfnm{Tamar}\binits{T.}},
  \bauthor{\bsnm{Peretz},~\bfnm{Ayelet}\binits{A.}},
  \bauthor{\bsnm{Magenheim},~\bfnm{Judith}\binits{J.}},
  \bauthor{\bsnm{Fogel},~\bfnm{Irit}\binits{I.}},
  \bauthor{\bsnm{Livneh},~\bfnm{Ayalah}\binits{A.}},
  \bauthor{\bsnm{Daitch},~\bfnm{Yutti}\binits{Y.}},
  \bauthor{\bsnm{Djian},~\bfnm{Esther~Oiknine}\binits{E.~O.}},
  \bauthor{\bsnm{Benedek},~\bfnm{Gil}\binits{G.}},
  \bauthor{\bsnm{Dor},~\bfnm{Yuval}\binits{Y.}},
  \bauthor{\bsnm{Wolf},~\bfnm{Dana~G.}\binits{D.~G.}} \AND
  \bauthor{\bsnm{Yassour},~\bfnm{Moran}\binits{M.}}
(\byear{2021}).
\btitle{{Lessons from applied large-scale pooling of 133,816 SARS-CoV-2 RT-PCR
  tests}}.
\bjournal{Science Translational Medicine}
\bvolume{13}
\bpages{1--8}.
\bdoi{10.1126/scitranslmed.abf2823}
\end{barticle}
\endbibitem

\bibitem[\protect\citeauthoryear{Bilder, Tebbs and
  Chen}{2010}]{bilder2010informative}
\begin{barticle}[author]
\bauthor{\bsnm{Bilder},~\bfnm{Christopher~R}\binits{C.~R.}},
  \bauthor{\bsnm{Tebbs},~\bfnm{Joshua~M}\binits{J.~M.}} \AND
  \bauthor{\bsnm{Chen},~\bfnm{Peng}\binits{P.}}
(\byear{2010}).
\btitle{Informative retesting}.
\bjournal{Journal of the American Statistical Association}
\bvolume{105}
\bpages{942--955}.
\end{barticle}
\endbibitem

\bibitem[\protect\citeauthoryear{Bilder and Tebbs}{2012}]{bilder2012pooled}
\begin{barticle}[author]
\bauthor{\bsnm{Bilder},~\bfnm{Christopher~R}\binits{C.~R.}} \AND
  \bauthor{\bsnm{Tebbs},~\bfnm{Joshua~M}\binits{J.~M.}}
(\byear{2012}).
\btitle{Pooled-testing procedures for screening high volume clinical specimens
  in heterogeneous populations}.
\bjournal{Statistics in medicine}
\bvolume{31}
\bpages{3261--3268}.
\end{barticle}
\endbibitem

\bibitem[\protect\citeauthoryear{Bilder, Tebbs and
  McMahan}{2019}]{bilder2019informative}
\begin{barticle}[author]
\bauthor{\bsnm{Bilder},~\bfnm{Christopher~R}\binits{C.~R.}},
  \bauthor{\bsnm{Tebbs},~\bfnm{Joshua~M}\binits{J.~M.}} \AND
  \bauthor{\bsnm{McMahan},~\bfnm{Christopher~S}\binits{C.~S.}}
(\byear{2019}).
\btitle{Informative group testing for multiplex assays}.
\bjournal{Biometrics}
\bvolume{75}
\bpages{278--288}.
\end{barticle}
\endbibitem

\bibitem[\protect\citeauthoryear{Cave}{2020}]{cave2020covid}
\begin{barticle}[author]
\bauthor{\bsnm{Cave},~\bfnm{Emma}\binits{E.}}
(\byear{2020}).
\btitle{COVID-19 super-spreaders: Definitional quandaries and implications}.
\bjournal{Asian Bioethics Review}
\bpages{1}.
\end{barticle}
\endbibitem

\bibitem[\protect\citeauthoryear{Chen, Tebbs and Bilder}{2009}]{chen2009group}
\begin{barticle}[author]
\bauthor{\bsnm{Chen},~\bfnm{Peng}\binits{P.}},
  \bauthor{\bsnm{Tebbs},~\bfnm{Joshua~M}\binits{J.~M.}} \AND
  \bauthor{\bsnm{Bilder},~\bfnm{Christopher~R}\binits{C.~R.}}
(\byear{2009}).
\btitle{Group testing regression models with fixed and random effects}.
\bjournal{Biometrics}
\bvolume{65}
\bpages{1270--1278}.
\end{barticle}
\endbibitem

\bibitem[\protect\citeauthoryear{Cirillo and Taleb}{2020}]{cirillo2020tail}
\begin{barticle}[author]
\bauthor{\bsnm{Cirillo},~\bfnm{Pasquale}\binits{P.}} \AND
  \bauthor{\bsnm{Taleb},~\bfnm{Nassim~Nicholas}\binits{N.~N.}}
(\byear{2020}).
\btitle{Tail risk of contagious diseases}.
\bjournal{Nature Physics}
\bvolume{16}
\bpages{606--613}.
\end{barticle}
\endbibitem

\bibitem[\protect\citeauthoryear{Collins}{2020}]{nyt}
\begin{bmisc}[author]
\bauthor{\bsnm{Collins},~\bfnm{Keith}\binits{K.}}
(\byear{2020}).
\btitle{Is Your State Doing Enough Coronavirus Testing?}
\bnote{[Online; posted 1-November-2020]}.
\end{bmisc}
\endbibitem

\bibitem[\protect\citeauthoryear{Cramer et~al.}{2021}]{cramer2021evaluation}
\begin{barticle}[author]
\bauthor{\bsnm{Cramer},~\bfnm{Estee~Y}\binits{E.~Y.}},
  \bauthor{\bsnm{Lopez},~\bfnm{Velma~K}\binits{V.~K.}},
  \bauthor{\bsnm{Niemi},~\bfnm{Jarad}\binits{J.}},
  \bauthor{\bsnm{George},~\bfnm{Glover~E}\binits{G.~E.}},
  \bauthor{\bsnm{Cegan},~\bfnm{Jeffrey~C}\binits{J.~C.}},
  \bauthor{\bsnm{Dettwiller},~\bfnm{Ian~D}\binits{I.~D.}},
  \bauthor{\bsnm{England},~\bfnm{William~P}\binits{W.~P.}},
  \bauthor{\bsnm{Farthing},~\bfnm{Matthew~W}\binits{M.~W.}},
  \bauthor{\bsnm{Hunter},~\bfnm{Robert~H}\binits{R.~H.}},
  \bauthor{\bsnm{Lafferty},~\bfnm{Brandon}\binits{B.}} \betal{et~al.}
(\byear{2021}).
\btitle{Evaluation of individual and ensemble probabilistic forecasts of
  COVID-19 mortality in the US}.
\bjournal{medRxiv}.
\end{barticle}
\endbibitem

\bibitem[\protect\citeauthoryear{Deckert, B{\"{a}}rnighausen and
  Kyei}{2020}]{Deckert2020SimulationTesting}
\begin{barticle}[author]
\bauthor{\bsnm{Deckert},~\bfnm{Andreas}\binits{A.}},
  \bauthor{\bsnm{B{\"{a}}rnighausen},~\bfnm{Till}\binits{T.}} \AND
  \bauthor{\bsnm{Kyei},~\bfnm{Nicholas N.~A.}\binits{N.~N.~A.}}
(\byear{2020}).
\btitle{{Simulation of pooled-sample analysis strategies for covid-19 mass
  testing}}.
\bjournal{Bulletin of the World Health Organization}
\bvolume{98}
\bpages{590--598}.
\bdoi{10.2471/BLT.20.257188}
\end{barticle}
\endbibitem

\bibitem[\protect\citeauthoryear{Dhillon et~al.}{2015}]{dhillon2015ebola}
\begin{barticle}[author]
\bauthor{\bsnm{Dhillon},~\bfnm{Ranu~S}\binits{R.~S.}},
  \bauthor{\bsnm{Srikrishna},~\bfnm{Devabhaktuni}\binits{D.}},
  \bauthor{\bsnm{Garry},~\bfnm{Robert~F}\binits{R.~F.}} \AND
  \bauthor{\bsnm{Chowell},~\bfnm{Gerardo}\binits{G.}}
(\byear{2015}).
\btitle{Ebola control: rapid diagnostic testing}.
\bjournal{The Lancet Infectious Diseases}
\bvolume{15}
\bpages{147--148}.
\end{barticle}
\endbibitem

\bibitem[\protect\citeauthoryear{Donnat and Holmes}{2021}]{donnat2021modeling}
\begin{barticle}[author]
\bauthor{\bsnm{Donnat},~\bfnm{Claire}\binits{C.}} \AND
  \bauthor{\bsnm{Holmes},~\bfnm{Susan}\binits{S.}}
(\byear{2021}).
\btitle{Modeling the heterogeneity in COVID-19's reproductive number and its
  impact on predictive scenarios}.
\bjournal{Journal of Applied Statistics}
\bvolume{0}
\bpages{1-29}.
\bdoi{10.1080/02664763.2021.1941806}
\end{barticle}
\endbibitem

\bibitem[\protect\citeauthoryear{Donnat et~al.}{2020}]{donnat2020bayesian}
\begin{binproceedings}[author]
\bauthor{\bsnm{Donnat},~\bfnm{Claire}\binits{C.}},
  \bauthor{\bsnm{Miolane},~\bfnm{Nina}\binits{N.}},
  \bauthor{\bsnm{Bunbury},~\bfnm{Freddy}\binits{F.}} \AND
  \bauthor{\bsnm{Kreindler},~\bfnm{Jack}\binits{J.}}
(\byear{2020}).
\btitle{A Bayesian Hierarchical Network for Combining Heterogeneous Data
  Sources in Medical Diagnoses}.
In \bbooktitle{Machine Learning for Health}
\bpages{53--84}.
\bpublisher{PMLR}.
\end{binproceedings}
\endbibitem

\bibitem[\protect\citeauthoryear{Donnat et~al.}{2021}]{donnat2021predictive}
\begin{barticle}[author]
\bauthor{\bsnm{Donnat},~\bfnm{Claire}\binits{C.}},
  \bauthor{\bsnm{Bunbury},~\bfnm{Freddy}\binits{F.}},
  \bauthor{\bsnm{Kreindler},~\bfnm{Jack}\binits{J.}},
  \bauthor{\bsnm{Filippidis},~\bfnm{Filippos~T}\binits{F.~T.}},
  \bauthor{\bsnm{El-Osta},~\bfnm{Austen}\binits{A.}},
  \bauthor{\bsnm{Esko},~\bfnm{T{\~o}nu}\binits{T.}} \AND
  \bauthor{\bsnm{Harris},~\bfnm{Matthew}\binits{M.}}
(\byear{2021}).
\btitle{A Predictive Modelling Framework for COVID-19 Transmission to Inform
  the Management of Mass Events}.
\bjournal{medRxiv}.
\end{barticle}
\endbibitem

\bibitem[\protect\citeauthoryear{Dorfman}{1943}]{Dorfman1943}
\begin{barticle}[author]
\bauthor{\bsnm{Dorfman},~\bfnm{Robert}\binits{R.}}
(\byear{1943}).
\btitle{{The Detection of Defective Members of Large Populations}}.
\bjournal{The Annals of Mathematical Statistics}
\bvolume{14}
\bpages{436--440}.
\end{barticle}
\endbibitem

\bibitem[\protect\citeauthoryear{Eunjung~Cha}{2021}]{wapo_schools}
\begin{bmisc}[author]
\bauthor{\bsnm{Eunjung~Cha},~\bfnm{Ariana}\binits{A.}}
(\byear{2021}).
\btitle{The future of coronavirus testing is in Greenville, N.C.}
\bnote{[Online; posted 28-April-2021]}.
\end{bmisc}
\endbibitem

\bibitem[\protect\citeauthoryear{{FDA}}{2020}]{FDA2020CoronavirusTesting}
\begin{bmisc}[author]
\bauthor{\bsnm{{FDA}}}
(\byear{2020}).
\btitle{{Coronavirus (COVID-19) Update: FDA Issues First Emergency
  Authorization for Sample Pooling in Diagnostic Testing}}.
\end{bmisc}
\endbibitem

\bibitem[\protect\citeauthoryear{Frayer}{2021}]{npr_india}
\begin{bmisc}[author]
\bauthor{\bsnm{Frayer},~\bfnm{Lauren}\binits{L.}}
(\byear{2021}).
\btitle{Oxygen Rationing, Test Shortages: India Caught Unprepared In COVID-19
  Crisis}.
\bnote{[Online; posted 24-April-2021]}.
\end{bmisc}
\endbibitem

\bibitem[\protect\citeauthoryear{Gandhi, Yokoe and
  Havlir}{2020}]{gandhi2020asymptomatic}
\begin{bmisc}[author]
\bauthor{\bsnm{Gandhi},~\bfnm{Monica}\binits{M.}},
  \bauthor{\bsnm{Yokoe},~\bfnm{Deborah~S}\binits{D.~S.}} \AND
  \bauthor{\bsnm{Havlir},~\bfnm{Diane~V}\binits{D.~V.}}
(\byear{2020}).
\btitle{Asymptomatic transmission, the Achilles? heel of current strategies to
  control COVID-19}.
\end{bmisc}
\endbibitem

\bibitem[\protect\citeauthoryear{Gastwirth}{2000}]{Gastwirth2000The2}
\begin{barticle}[author]
\bauthor{\bsnm{Gastwirth},~\bfnm{J.~L.}\binits{J.~L.}}
(\byear{2000}).
\btitle{{The efficiency of pooling in the detection of rare mutations [2]}}.
\bjournal{American Journal of Human Genetics}
\bvolume{67}
\bpages{1036--1039}.
\bdoi{10.1086/303097}
\end{barticle}
\endbibitem

\bibitem[\protect\citeauthoryear{Gaydos}{2005}]{gaydos2005nucleic}
\begin{barticle}[author]
\bauthor{\bsnm{Gaydos},~\bfnm{Charlotte~A}\binits{C.~A.}}
(\byear{2005}).
\btitle{Nucleic acid amplification tests for gonorrhea and chlamydia: practice
  and applications}.
\bjournal{Infectious Disease Clinics}
\bvolume{19}
\bpages{367--386}.
\end{barticle}
\endbibitem

\bibitem[\protect\citeauthoryear{G{\'o}mez-Carballa
  et~al.}{2020}]{gomez2020mapping}
\begin{barticle}[author]
\bauthor{\bsnm{G{\'o}mez-Carballa},~\bfnm{Alberto}\binits{A.}},
  \bauthor{\bsnm{Bello},~\bfnm{Xabier}\binits{X.}},
  \bauthor{\bsnm{Pardo-Seco},~\bfnm{Jacobo}\binits{J.}},
  \bauthor{\bsnm{Martin{\'o}n-Torres},~\bfnm{Federico}\binits{F.}} \AND
  \bauthor{\bsnm{Salas},~\bfnm{Antonio}\binits{A.}}
(\byear{2020}).
\btitle{Mapping genome variation of SARS-CoV-2 worldwide highlights the impact
  of COVID-19 super-spreaders}.
\bjournal{Genome Research}
\bvolume{30}
\bpages{1434--1448}.
\end{barticle}
\endbibitem

\bibitem[\protect\citeauthoryear{He et~al.}{2020}]{he2020proportion}
\begin{barticle}[author]
\bauthor{\bsnm{He},~\bfnm{Jingjing}\binits{J.}},
  \bauthor{\bsnm{Guo},~\bfnm{Yifei}\binits{Y.}},
  \bauthor{\bsnm{Mao},~\bfnm{Richeng}\binits{R.}} \AND
  \bauthor{\bsnm{Zhang},~\bfnm{Jiming}\binits{J.}}
(\byear{2020}).
\btitle{Proportion of asymptomatic coronavirus disease 2019: A systematic
  review and meta-analysis}.
\bjournal{Journal of medical virology}.
\end{barticle}
\endbibitem

\bibitem[\protect\citeauthoryear{Kim et~al.}{2007}]{Kim2007ComparisonError}
\begin{barticle}[author]
\bauthor{\bsnm{Kim},~\bfnm{Hae~Young}\binits{H.~Y.}},
  \bauthor{\bsnm{Hudgens},~\bfnm{Michael~G.}\binits{M.~G.}},
  \bauthor{\bsnm{Dreyfuss},~\bfnm{Jonathan~M.}\binits{J.~M.}},
  \bauthor{\bsnm{Westreich},~\bfnm{Daniel~J.}\binits{D.~J.}} \AND
  \bauthor{\bsnm{Pilcher},~\bfnm{Christopher~D.}\binits{C.~D.}}
(\byear{2007}).
\btitle{{Comparison of group testing algorithms for case identification in the
  presence of test error}}.
\bjournal{Biometrics}
\bvolume{63}
\bpages{1152--1163}.
\bdoi{10.1111/j.1541-0420.2007.00817.x}
\end{barticle}
\endbibitem

\bibitem[\protect\citeauthoryear{Koh et~al.}{2020}]{Koh2020}
\begin{barticle}[author]
\bauthor{\bsnm{Koh},~\bfnm{Wee~Chian}\binits{W.~C.}},
  \bauthor{\bsnm{Naing},~\bfnm{Lin}\binits{L.}},
  \bauthor{\bsnm{Chaw},~\bfnm{Liling}\binits{L.}},
  \bauthor{\bsnm{Rosledzana},~\bfnm{Muhammad~Ali}\binits{M.~A.}},
  \bauthor{\bsnm{Alikhan},~\bfnm{Mohammad~Fathi}\binits{M.~F.}},
  \bauthor{\bsnm{Jamaludin},~\bfnm{Sirajul~Adli}\binits{S.~A.}},
  \bauthor{\bsnm{Amin},~\bfnm{Faezah}\binits{F.}},
  \bauthor{\bsnm{Omar},~\bfnm{Asiah}\binits{A.}},
  \bauthor{\bsnm{Shazli},~\bfnm{Alia}\binits{A.}},
  \bauthor{\bsnm{Griffith},~\bfnm{Matthew}\binits{M.}},
  \bauthor{\bsnm{Pastore},~\bfnm{Roberta}\binits{R.}} \AND
  \bauthor{\bsnm{Wong},~\bfnm{Justin}\binits{J.}}
(\byear{2020}).
\btitle{{What do we know about SARS-CoV-2 transmission? A systematic review and
  meta-analysis of the secondary attack rate and associated risk factors}}.
\bjournal{PLoS ONE}
\bvolume{15}
\bpages{1--23}.
\bdoi{10.1371/journal.pone.0240205}
\end{barticle}
\endbibitem

\bibitem[\protect\citeauthoryear{Larremore
  et~al.}{2020}]{Larremore2020TestScreening.}
\begin{barticle}[author]
\bauthor{\bsnm{Larremore},~\bfnm{Daniel~B}\binits{D.~B.}},
  \bauthor{\bsnm{Wilder},~\bfnm{Bryan}\binits{B.}},
  \bauthor{\bsnm{Lester},~\bfnm{Evan}\binits{E.}},
  \bauthor{\bsnm{Shehata},~\bfnm{Soraya}\binits{S.}},
  \bauthor{\bsnm{Burke},~\bfnm{James~M}\binits{J.~M.}},
  \bauthor{\bsnm{Hay},~\bfnm{James~A}\binits{J.~A.}},
  \bauthor{\bsnm{Tambe},~\bfnm{Milind}\binits{M.}},
  \bauthor{\bsnm{Mina},~\bfnm{Michael~J}\binits{M.~J.}} \AND
  \bauthor{\bsnm{Parker},~\bfnm{Roy}\binits{R.}}
(\byear{2020}).
\btitle{{Test sensitivity is secondary to frequency and turnaround time for
  COVID-19 screening.}}
\bjournal{Science advances}.
\bdoi{10.1126/sciadv.abd5393}
\end{barticle}
\endbibitem

\bibitem[\protect\citeauthoryear{Lin et~al.}{2020}]{Lin2020PositivelyCosts}
\begin{barticle}[author]
\bauthor{\bsnm{Lin},~\bfnm{Yi-Jheng}\binits{Y.-J.}},
  \bauthor{\bsnm{Yu},~\bfnm{Che-Hao}\binits{C.-H.}},
  \bauthor{\bsnm{Liu},~\bfnm{Tzu-Hsuan}\binits{T.-H.}},
  \bauthor{\bsnm{Chang},~\bfnm{Cheng-Shang}\binits{C.-S.}} \AND
  \bauthor{\bsnm{Chen},~\bfnm{Wen-Tsuen}\binits{W.-T.}}
(\byear{2020}).
\btitle{{Positively Correlated Samples Save Pooled Testing Costs}}.
\end{barticle}
\endbibitem

\bibitem[\protect\citeauthoryear{McMahan, Tebbs and
  Bilder}{2012a}]{McMahan2012InformativeScreening}
\begin{barticle}[author]
\bauthor{\bsnm{McMahan},~\bfnm{Christopher~S.}\binits{C.~S.}},
  \bauthor{\bsnm{Tebbs},~\bfnm{Joshua~M.}\binits{J.~M.}} \AND
  \bauthor{\bsnm{Bilder},~\bfnm{Christopher~R.}\binits{C.~R.}}
(\byear{2012}a).
\btitle{{Informative Dorfman Screening}}.
\bjournal{Biometrics}
\bvolume{68}
\bpages{287--296}.
\bdoi{10.1111/j.1541-0420.2011.01644.x}
\end{barticle}
\endbibitem

\bibitem[\protect\citeauthoryear{McMahan, Tebbs and
  Bilder}{2012b}]{mcmahan2012two}
\begin{barticle}[author]
\bauthor{\bsnm{McMahan},~\bfnm{Christopher~S}\binits{C.~S.}},
  \bauthor{\bsnm{Tebbs},~\bfnm{Joshua~M}\binits{J.~M.}} \AND
  \bauthor{\bsnm{Bilder},~\bfnm{Christopher~R}\binits{C.~R.}}
(\byear{2012}b).
\btitle{Two-dimensional informative array testing}.
\bjournal{Biometrics}
\bvolume{68}
\bpages{793--804}.
\end{barticle}
\endbibitem

\bibitem[\protect\citeauthoryear{Mina, Parker and
  Larremore}{2020}]{doi:10.1056/NEJMp2025631}
\begin{barticle}[author]
\bauthor{\bsnm{Mina},~\bfnm{Michael~J.}\binits{M.~J.}},
  \bauthor{\bsnm{Parker},~\bfnm{Roy}\binits{R.}} \AND
  \bauthor{\bsnm{Larremore},~\bfnm{Daniel~B.}\binits{D.~B.}}
(\byear{2020}).
\btitle{Rethinking Covid-19 Test Sensitivity — A Strategy for Containment}.
\bjournal{New England Journal of Medicine}
\bvolume{383}
\bpages{e120}.
\bdoi{10.1056/NEJMp2025631}
\end{barticle}
\endbibitem

\bibitem[\protect\citeauthoryear{Mwai}{2021}]{bbc_africa}
\begin{bmisc}[author]
\bauthor{\bsnm{Mwai},~\bfnm{Peter}\binits{P.}}
(\byear{2021}).
\btitle{Coronavirus in Africa: Concern growing over third wave of Covid-19
  infections}.
\bnote{[Online; posted 7-June-2021]}.
\end{bmisc}
\endbibitem

\bibitem[\protect\citeauthoryear{Nouvellet et~al.}{2015}]{nouvellet2015role}
\begin{barticle}[author]
\bauthor{\bsnm{Nouvellet},~\bfnm{Pierre}\binits{P.}},
  \bauthor{\bsnm{Garske},~\bfnm{Tini}\binits{T.}},
  \bauthor{\bsnm{Mills},~\bfnm{Harriet~L}\binits{H.~L.}},
  \bauthor{\bsnm{Nedjati-Gilani},~\bfnm{Gemma}\binits{G.}},
  \bauthor{\bsnm{Hinsley},~\bfnm{Wes}\binits{W.}},
  \bauthor{\bsnm{Blake},~\bfnm{Isobel~M}\binits{I.~M.}},
  \bauthor{\bsnm{Van~Kerkhove},~\bfnm{Maria~D}\binits{M.~D.}},
  \bauthor{\bsnm{Cori},~\bfnm{Anne}\binits{A.}},
  \bauthor{\bsnm{Dorigatti},~\bfnm{Ilaria}\binits{I.}},
  \bauthor{\bsnm{Jombart},~\bfnm{Thibaut}\binits{T.}} \betal{et~al.}
(\byear{2015}).
\btitle{The role of rapid diagnostics in managing Ebola epidemics}.
\bjournal{Nature}
\bvolume{528}
\bpages{S109--S116}.
\end{barticle}
\endbibitem

\bibitem[\protect\citeauthoryear{Oran and Topol}{2020}]{oran2020prevalence}
\begin{barticle}[author]
\bauthor{\bsnm{Oran},~\bfnm{Daniel~P}\binits{D.~P.}} \AND
  \bauthor{\bsnm{Topol},~\bfnm{Eric~J}\binits{E.~J.}}
(\byear{2020}).
\btitle{Prevalence of Asymptomatic SARS-CoV-2 Infection: A Narrative Review}.
\bjournal{Annals of Internal Medicine}.
\end{barticle}
\endbibitem

\bibitem[\protect\citeauthoryear{Pollock and
  Lancaster}{2020}]{pollock2020asymptomatic}
\begin{barticle}[author]
\bauthor{\bsnm{Pollock},~\bfnm{Allyson~M.}\binits{A.~M.}} \AND
  \bauthor{\bsnm{Lancaster},~\bfnm{James}\binits{J.}}
(\byear{2020}).
\btitle{{Asymptomatic transmission of covid-19}}.
\bjournal{BMJ}
\bvolume{371}
\bpages{m4851}.
\bdoi{10.1136/bmj.m4851}
\end{barticle}
\endbibitem

\bibitem[\protect\citeauthoryear{Rannan-Eliya
  et~al.}{2021}]{rannan2021increased}
\begin{barticle}[author]
\bauthor{\bsnm{Rannan-Eliya},~\bfnm{Ravindra~Prasan}\binits{R.~P.}},
  \bauthor{\bsnm{Wijemunige},~\bfnm{Nilmini}\binits{N.}},
  \bauthor{\bsnm{Gunawardana},~\bfnm{JRNA}\binits{J.}},
  \bauthor{\bsnm{Amarasinghe},~\bfnm{Sarasi~N}\binits{S.~N.}},
  \bauthor{\bsnm{Sivagnanam},~\bfnm{Ishwari}\binits{I.}},
  \bauthor{\bsnm{Fonseka},~\bfnm{Sachini}\binits{S.}},
  \bauthor{\bsnm{Kapuge},~\bfnm{Yasodhara}\binits{Y.}} \AND
  \bauthor{\bsnm{Sigera},~\bfnm{Chathurani~P}\binits{C.~P.}}
(\byear{2021}).
\btitle{Increased Intensity Of PCR Testing Reduced COVID-19 Transmission Within
  Countries During The First Pandemic Wave: Study examines increased intensity
  of reverse transcription--polymerase chain reaction (PCR) testing and its
  impact on COVID-19 transmission.}
\bjournal{Health Affairs}
\bpages{10--1377}.
\end{barticle}
\endbibitem

\bibitem[\protect\citeauthoryear{Rewley}{2020}]{Rewley2020SpecimenStudy}
\begin{barticle}[author]
\bauthor{\bsnm{Rewley},~\bfnm{Jeffrey}\binits{J.}}
(\byear{2020}).
\btitle{{Specimen pooling to conserve additional testing resources when
  persons’ infection status is correlated: A simulation study}}.
\bjournal{Epidemiology}
\bvolume{31}
\bpages{832--835}.
\bdoi{10.1097/EDE.0000000000001244}
\end{barticle}
\endbibitem

\bibitem[\protect\citeauthoryear{Stevens et~al.}{2021}]{harvard2021}
\begin{barticle}[author]
\bauthor{\bsnm{Stevens},~\bfnm{Jennifer~P.}\binits{J.~P.}},
  \bauthor{\bsnm{Horng},~\bfnm{Steven}\binits{S.}},
  \bauthor{\bsnm{O’Donoghue},~\bfnm{Ashley}\binits{A.}},
  \bauthor{\bsnm{Moravick},~\bfnm{Sarah}\binits{S.}} \AND
  \bauthor{\bsnm{Weiss},~\bfnm{Anthony}\binits{A.}}
(\byear{2021}).
\btitle{How One Boston Hospital Built a Covid-19 Forecasting System}.
\bjournal{Harvard Business Review}.
\end{barticle}
\endbibitem

\bibitem[\protect\citeauthoryear{Tom and Mina}{2020}]{tom2020interpret}
\begin{barticle}[author]
\bauthor{\bsnm{Tom},~\bfnm{Michael~R}\binits{M.~R.}} \AND
  \bauthor{\bsnm{Mina},~\bfnm{Michael~J}\binits{M.~J.}}
(\byear{2020}).
\btitle{To interpret the SARS-CoV-2 test, consider the cycle threshold value}.
\bjournal{Clinical Infectious Diseases}.
\end{barticle}
\endbibitem

\bibitem[\protect\citeauthoryear{Tso et~al.}{2021}]{Tso2021}
\begin{barticle}[author]
\bauthor{\bsnm{Tso},~\bfnm{Chak~Foon}\binits{C.~F.}},
  \bauthor{\bsnm{Garikipati},~\bfnm{Anurag}\binits{A.}},
  \bauthor{\bsnm{Green-Saxena},~\bfnm{Abigail}\binits{A.}},
  \bauthor{\bsnm{Mao},~\bfnm{Qingqing}\binits{Q.}} \AND
  \bauthor{\bsnm{Das},~\bfnm{Ritankar}\binits{R.}}
(\byear{2021}).
\btitle{{Correlation of population SARS-CoV-2 cycle threshold values to local
  disease dynamics: Exploratory observational study}}.
\bjournal{JMIR Public Health and Surveillance}
\bvolume{7}.
\bdoi{10.2196/28265}
\end{barticle}
\endbibitem

\bibitem[\protect\citeauthoryear{Tu, Litvak and
  Pagano}{1995}]{tu1995informativeness}
\begin{barticle}[author]
\bauthor{\bsnm{Tu},~\bfnm{Xin~Ming}\binits{X.~M.}},
  \bauthor{\bsnm{Litvak},~\bfnm{Eugene}\binits{E.}} \AND
  \bauthor{\bsnm{Pagano},~\bfnm{Marcello}\binits{M.}}
(\byear{1995}).
\btitle{On the informativeness and accuracy of pooled testing in estimating
  prevalence of a rare disease: application to HIV screening}.
\bjournal{Biometrika}
\bvolume{82}
\bpages{287--297}.
\end{barticle}
\endbibitem

\bibitem[\protect\citeauthoryear{Wang et~al.}{2021}]{Wang2021}
\begin{barticle}[author]
\bauthor{\bsnm{Wang},~\bfnm{Hannah}\binits{H.}},
  \bauthor{\bsnm{Hogan},~\bfnm{Catherine~A.}\binits{C.~A.}},
  \bauthor{\bsnm{Miller},~\bfnm{Jacob~A.}\binits{J.~A.}},
  \bauthor{\bsnm{Sahoo},~\bfnm{Malaya~K.}\binits{M.~K.}},
  \bauthor{\bsnm{Huang},~\bfnm{Chun~Hong}\binits{C.~H.}},
  \bauthor{\bsnm{Mfuh},~\bfnm{Kenji~O.}\binits{K.~O.}},
  \bauthor{\bsnm{Sibai},~\bfnm{Mamdouh}\binits{M.}},
  \bauthor{\bsnm{Zehnder},~\bfnm{James}\binits{J.}},
  \bauthor{\bsnm{Hickey},~\bfnm{Brendan}\binits{B.}},
  \bauthor{\bsnm{Sinnott-Armstrong},~\bfnm{Nasa}\binits{N.}} \AND
  \bauthor{\bsnm{Pinsky},~\bfnm{Benjamin~A.}\binits{B.~A.}}
(\byear{2021}).
\btitle{{Performance of nucleic acid amplification tests for detection of
  severe acute respiratory syndrome coronavirus 2 in prospectively pooled
  specimens}}.
\bjournal{Emerging Infectious Diseases}
\bvolume{27}
\bpages{92--103}.
\bdoi{10.3201/eid2701.203379}
\end{barticle}
\endbibitem

\bibitem[\protect\citeauthoryear{Wein and Zenios}{1996}]{wein1996pooled}
\begin{barticle}[author]
\bauthor{\bsnm{Wein},~\bfnm{Lawrence~M}\binits{L.~M.}} \AND
  \bauthor{\bsnm{Zenios},~\bfnm{Stefanos~A}\binits{S.~A.}}
(\byear{1996}).
\btitle{Pooled testing for HIV screening: capturing the dilution effect}.
\bjournal{Operations Research}
\bvolume{44}
\bpages{543--569}.
\end{barticle}
\endbibitem

\bibitem[\protect\citeauthoryear{Yamamura and
  Hino}{2007}]{yamamura2007estimation}
\begin{barticle}[author]
\bauthor{\bsnm{Yamamura},~\bfnm{Kohji}\binits{K.}} \AND
  \bauthor{\bsnm{Hino},~\bfnm{Akihiro}\binits{A.}}
(\byear{2007}).
\btitle{Estimation of the proportion of defective units by using group testing
  under the existence of a threshold of detection}.
\bjournal{Communications in Statistics—Simulation and Computation}
\bvolume{36}
\bpages{949--957}.
\end{barticle}
\endbibitem

\bibitem[\protect\citeauthoryear{Zhang
  et~al.}{2020a}]{Zhang2020FamilialAsymptomatic}
\begin{barticle}[author]
\bauthor{\bsnm{Zhang},~\bfnm{Jinjun}\binits{J.}},
  \bauthor{\bsnm{Tian},~\bfnm{Sijia}\binits{S.}},
  \bauthor{\bsnm{Lou},~\bfnm{Jing}\binits{J.}} \AND
  \bauthor{\bsnm{Chen},~\bfnm{Yuguo}\binits{Y.}}
(\byear{2020}a).
\btitle{{Familial cluster of COVID-19 infection from an asymptomatic}}.
\bjournal{Critical Care}
\bvolume{24}
\bpages{7--9}.
\bdoi{10.1186/s13054-020-2817-7}
\end{barticle}
\endbibitem

\bibitem[\protect\citeauthoryear{Zhang et~al.}{2020b}]{zhang2020evaluating}
\begin{barticle}[author]
\bauthor{\bsnm{Zhang},~\bfnm{Yunjun}\binits{Y.}},
  \bauthor{\bsnm{Li},~\bfnm{Yuying}\binits{Y.}},
  \bauthor{\bsnm{Wang},~\bfnm{Lu}\binits{L.}},
  \bauthor{\bsnm{Li},~\bfnm{Mingyuan}\binits{M.}} \AND
  \bauthor{\bsnm{Zhou},~\bfnm{Xiaohua}\binits{X.}}
(\byear{2020}b).
\btitle{Evaluating transmission heterogeneity and super-spreading event of
  COVID-19 in a metropolis of China}.
\bjournal{International Journal of Environmental Research and Public Health}
\bvolume{17}
\bpages{3705}.
\end{barticle}
\endbibitem

\bibitem[\protect\citeauthoryear{Zhou et~al.}{2020}]{zhou2020spatiotemporal}
\begin{barticle}[author]
\bauthor{\bsnm{Zhou},~\bfnm{Yiwang}\binits{Y.}},
  \bauthor{\bsnm{Wang},~\bfnm{Lili}\binits{L.}},
  \bauthor{\bsnm{Zhang},~\bfnm{Leyao}\binits{L.}},
  \bauthor{\bsnm{Shi},~\bfnm{Lan}\binits{L.}},
  \bauthor{\bsnm{Yang},~\bfnm{Kangping}\binits{K.}},
  \bauthor{\bsnm{He},~\bfnm{Jie}\binits{J.}},
  \bauthor{\bsnm{Zhao},~\bfnm{Bangyao}\binits{B.}},
  \bauthor{\bsnm{Overton},~\bfnm{William}\binits{W.}},
  \bauthor{\bsnm{Purkayastha},~\bfnm{Soumik}\binits{S.}} \AND
  \bauthor{\bsnm{Song},~\bfnm{Peter}\binits{P.}}
(\byear{2020}).
\btitle{A spatiotemporal epidemiological prediction model to inform
  county-level COVID-19 risk in the United States}.
\bjournal{Harvard Data Science Review}.
\end{barticle}
\endbibitem

\end{thebibliography}
\pagebreak 
\appendix
\counterwithin{figure}{section}

\section{Mathematical Framework}\label{sec:appendix_comp}
\subsection{Analysing the effect of correlations on the $C_t$ values}
In this appendix, we provide a more detailed discussion on the effects of the dilution under correlations on the pooled sample's $C_t$ value. As highlighted in the main document, this quantity is indeed key in establishing the expected sensitivity and efficiency of the procedure.

Consistently with the main text, let us denote as $ C_{t}^{\text{dilution}}$ the $C_t$ value of the dilution, and by $C_t^{(i)}$ the $C_t$ value of the individual test for sample $i$.  
Let us consider a case where $K$ samples are positive (which, without loss of generality, we can take to be the first $K$, and which we rank by increasing $C_t$, so that $C_t^{(1)}< C_t^{(2)}< \cdots < C_t^{(K)}$). The $C_t$ sample of the dilution can thus be written as:
  \begin{equation}
  \label{eq:ct_dilution_change1}
    \begin{split}
         C_{t}^{\text{dilution}} & \overset{D}{=}  - \log_2 \left( {\sum_{i=1}^K  2^{-C_{t}^{(i)}}} \right) +\log_2(n) \overset{D}{=} - \log_2 \left( 2^{-C_{t}^{(1)}} {\sum_{i=1}^k  2^{C_{t}^{(1)} -C_{t}^{(i)}}} \right) +\log_2(n) \\
         & \overset{D}{=} C_{t}^{(1)} + \log_2(n) - \log_2 \left( 1 + \sum_{i=2}^K 2^{\underbrace{C_t^{(1)}-C_{t}^{(i)}}_{\leq0}} \right) \text{ \tiny since $C_t^{(1)}< C_t^{(j)}, \forall j \geq 2$} \\
           & \overset{D}{=}  \min_{i\leq k}(C_{t}^{(i)}) + \log_2(n) - \frac{1}{\log(2)} \sum_{i=2}^K   2^{C_{t}^{(1)}-C_{t}^{(i)}} + o \left(\frac{1}{\log(2)} \sum_{i=2}^k   2^{C_{t}^{(1)}-C_{t}^{(i)}} \right) \\ 
    \end{split}
\end{equation} 
where the last line follows by Taylor-expansion of the $\log$ function around 1. This shows that, as described in the main text, the behavior of the $C_t$ value is dominated by the behavior of the minimum $C_t$ of $k$ samples, with an offset value of $\log_2(n)$. This also allows us to better understand the behavior of the pooled sensitivity, and characterize regimes where we can expect the grouping to improve upon the individual testing one through the hitchhiker effect. 

To make this statement more explicit, we note that the real data that we have at hand can be well approximated by a Weibull distribution with shape parameter $k = 4.55$ and scale $\lambda=29.86$. As stated in the main text, while this data was collected at a specific site, and the exact values of the parameters might not generalize to other population, we expect that the shape of the distribution will stay the same, but that the moments will be shifted. Thus, the equations that we derive, and which rely on assuming a Weibull distribution can serve as a good first indication, and can be adapted to any population of interest  by replacing the shape and scale values by the appropriate ones.  The quantiles of the Weibull distribution that approximates the distribution of the individual $C_t$s are indeed given by the function:
$$  F(x) =1-e^{-(x/\lambda)^k}$$
We also know that the mean is:  $\mathbb{E}[C_t^{(i)}] = \lambda \Gamma(1+\frac{1}{k})$.
Thus, approximating  $ C_{t}^{\text{dilution}} $ by $\min_{i\leq k}(C_{t}^{(i)}) + \log_2(n)$, the probability that the pooled $C_t$ of $K$  positive samples to worse (i.e, higher) than that of the average for individual testing can be expressed as:
\begin{equation}\label{eq:ct}
    \begin{split}
   \mathbb{P} [ \min_{i \leq K} (C_{t}^{(i)}) + \log_2(n) > \mathbb{E}[C_t^{(i)} ]   ] & =  \prod_{i=1}^K   \mathbb{P} [ C_{t}^{(i)} > \mathbb{E}[C_t^{(i)} ] - \log_2(n)] \quad  \text{\tiny  by independence of the $C_t$ values} \\
    & =   e^{-K (\frac{\mathbb{E}[C_t^{(i)} ] - \log_2(n)}{\lambda})^{k}}  =   e^{-K (\frac{ \lambda \Gamma(1 + \frac{1}{k}) - \log_2(n)}{\lambda})^{k}}  \\
   &\leq    e^{-K \Gamma(1 + \frac{1}{k})^{k}} \Big(1 - \frac{\log_2(n)}{\lambda} kK \Gamma(1 + \frac{1}{k})^{k-1}  \\
   &+  ( k^2K^2 \Gamma(1 + \frac{1}{k})^{2k-2}  - (k-1) kK \Gamma(1 + \frac{1}{k})^{k-2}  ) \frac{\log_2(n)^2}{\lambda^2}   \Big)\\
   &\leq    e^{-K \Gamma(1 + \frac{1}{k})^{k}} \Big(1 -  0.11 K  \log_2(n) +  ( 0.013 K^2  -  0.014 K) \log_2(n)^2   \Big)\\
   &\leq  0.52^{K}  (1 + 0.11 K \log_2(n))   \\
    \end{split}
\end{equation}

There is thus an exponential decay of this probability with the number of positive samples in the pool. 

We can similarly quantify the probability that the $C_t$ of the grouped sample will exceed that of the individual sample:
\begin{equation}
    \begin{split}
   \mathbb{P} [ \min_K(C_{t}^{(i)}) + \log_2(n) >  C_t^{(1)}  ] & =   \mathbb{P} [ C_{t}^{(1)} > C_t^{(2)} - \log_2(n)]^K \quad  \text{\tiny  by independence of the $C_t$ values} \\
   & = \mathbb{E}[ e^{ -( \frac{X-\log_2(n)}{\lambda})^k}] ^{K}   \approx \mathbb{E}[ e^{ -( \frac{X}{\lambda})^k} (1 +  \frac{\log_2(n)}{\lambda} k \frac{X^{k-1}}{\lambda^k} )] ^{K}  \\\
        &\leq \Big(\int_{x\geq 0}  ( 1   + k\frac{\log_2(n)}{\lambda^2} (\frac{x}{\lambda})^{k-1} )  \frac{k}{\lambda} (\frac{x}{\lambda})^{k-1} e^{-2(x/\lambda)^k}dx \Big)^{K}  \\
        & \approx \Big(\frac{1}{2}  + k\frac{\log_2(n)}{2 \lambda^{k+1}}  \mathbb{E}_{k, \lambda/2^{1/k}}[X^{k-1}] \Big)^{K}  \\
         & \approx\Big(\frac{1}{2}  + k\frac{\log_2(n)}{2 \lambda^{k+1}} (\frac{\lambda}{2^{1/k}})^{4} \Gamma(1+ \frac{4}{k}] \Big)^{K}  \text{\footnotesize  since $x \to \Gamma(x)$ is increasing on $[1, \infty)$}\\
         & \approx \Big(\frac{1}{2}  + 1.0 \times  10^{-2}{\log_2(n)}] \Big)^{K}  \\
    \end{split}\label{eq:ct2}
\end{equation}



The previous equations stem from the fact that the $C_t$ values of infected samples can be considered as independent: no evidence has linked so far the $C_t$ value of a sample to either individual characteristics or virus variants. As such, in our framework, while the probability of samples being positive within the pool are not independent (due to a potential network effect), the $C_t$ values themselves of the infected samples are. This also shows that the largest benefit occurs when going from $1$ to 2 contaminated samples: the probability that the $C_t$ dilution of the pooled samples is greater than that of the original samples becomes roughly $\frac{1}{2^2} + \frac{2\log_2(n)}{100}$, which, for the size of pools that we are considering (less than 100), is always less than 50\%, and by Eq.\ref{eq:ct}, the probability that the $C_t$ value of the mixture is greater than the expected $C_t$ of the individual samples is less than $0.67$\% for $K=2$, and 0.45 for $K=3$ (assuming worse case $n=100$). Thus, after $K=2$ to 3 positive samples, the hitchhiker effect ensures that the $C_t$ value of the mixture will be favorable.

\subsection{Probabilities}

Having established the behavior of the $C_t$ of the dilution as a function of the number of positive samples $K$, we turn to the evalution of the law of $K$ as a function of community $\pi$ and network transmission $\tau$. 

\xhdr{Elementary properties} We begin by listing a few elementary properties of these distributions, as summarized in Table~\ref{tab:properties} of the main text. Assuming that the prevalence $\pi$ is fixed and known, then the probability ($p$) of a given individual $i$ being infected is:

\begin{equation}
\label{eq:indiv_prob}
    \begin{split}
        p= \mathbb{P}[Y_i=1] &= \mathbb{P}[T_i^{(o)}] + \mathbb{P}[(1-T_i^{(o)})T_i^{(i)}]\\
        &= \pi + (1-\pi) \left( \mathbb{P} \left[ \left( \sum_{j\neq i} Y_j \right) \geq 1, T_i^{(i)} = 1 \right] \right) \\
        &= \pi + (1-\pi)\sum_{k=1}^{n-1} \left( \mathbb{P} \left[ \left( \sum_{j\neq i} Y_j \right) =k, T_i^{(i)} = 1 \right] \right) \\
        &= \pi + (1-\pi)\sum_{k=1}^{n-1} \left( {n-1 \choose k} \pi^k (1-\pi)^{n-1-k} (1- (1-\tau)^k) \right)\\
                &= \pi + (1-\pi)\sum_{k=1}^{n-1} \left( {n-1 \choose k} \pi^k (1-\pi)^{n-1-k}  - {n-1 \choose k} ((1-\tau) \pi)^k(1-\pi)^{n-1-k} \right)  \\
&= \pi + (1-\pi) \left( 1  - (1-\pi)^{n-1} - (\pi(1-\tau) + (1-\pi))^{n-1}  + (1-\pi)^{n-1} \right)  \\
&= \pi + (1-\pi) \left( 1  - (1-\pi\tau)^{n-1}   \right)  \\
    &= 1-(1-\pi)(1-\pi\tau)^{n-1}\\
&=1-(1-\pi)(1-(n-1)\pi\tau + o(\pi\tau))  = (1+(n-1)\tau)\pi + o(\tau \pi)
    \end{split}
\end{equation}

The covariance with other variables is thus:
\begin{equation}
\label{eq:covariance}
    \begin{split}
       \text{Cov}[Y_i, Y_j] &= \mathbb{E}[1\{Y_i=1\}1\{Y_j=1\}]  - p^2\\
       &= \pi^2 + 2\pi(1-\pi)\sum_{k=0}^{n-2} {n-2 \choose k} \pi^k (1-\pi)^{n-2-k} (1- (1-\tau)^{k+1}) \\
       &+   (1-\pi)^2 \sum_{k=1}^{n-2} {n-2 \choose k} \pi^k (1-\pi)^{n-2-k} (1- (1-\tau)^k)^2 - p^2\\
       &= \pi^2 + 2\pi(1-\pi)\big[1-(1-\tau)(1-\pi\tau)^{n-2}\big]\\
       &+   (1-\pi)^2 ( 1 + (1-\pi + (1-\tau)^2 \pi  )^{n-2} - 2(1-\pi \tau)^{n-2}) -p^2\\
       &= 1 - 2(1-\pi\tau)^{n-1} (1-\pi)  +(1-\pi)^2(1-\pi + (1-\tau)^2 \pi  )^{n-2}\\
       &- ( 1 + (1-\pi)^2(1-\pi\tau)^{2n-2} -2(1-\pi)(1-\pi\tau)^{n-1})\\
   &=  (1-\pi)^2\Big[(1-\pi + (1-\tau)^2 \pi  )^{n-2}-  (1-\pi\tau)^{2n-2}\Big]\\ 
   &= \pi \tau( 2 + (n-2)\tau) + o(\tau \pi)\\
       \end{split}
\end{equation}
Hence:
$$\rho_{ij} = \frac{\pi \tau( 1 - \tau + (1 + (n-1)\tau)) + o(\pi)}{(1+(n-1)\tau)\pi  + o(\tau \pi)} = (1+\frac{1-\tau}{1+(n-1)\tau})\tau + o(\tau)$$
This expression shows that the correlation is positive, and increasing as a function of $\tau$.

\xhdr{Effect on the average number of positive samples} We now turn to the impact of correlations on the expected number of positive samples. Consistently with the notation adopted in the main text, we denote $S = \sum_{i=1}^n Y_i$ the total number of positive samples.

\underline{$H_0$: Under the Null}: $S = \sum_{i=1}^n T_i^{\text{(community)}} \sim \text{Binom}(n, \pi) $, and the mean number of infected people is thus $n\pi$. Note that in the regimes that we are considering in this paper, $ n\pi \leq 0.10$ (Assumption 1), so that the mode of the distribution is in fact $0$. Even among positive pools, the mode of the number of infected samples is simply $  \frac{\lfloor (n + 1) \pi  \rfloor  }{1-(1-\pi)^n} =  \frac{\lfloor (n + 1) \pi  \rfloor  }{ n \pi + o(n \pi )} $,  We also note that:
\begin{equation}
    \begin{split}
    \mathbb{P}[S \geq 2| S \geq 1 ]  & = \frac{1-(1-\pi)^n - n\pi(1-\pi)^{n-1}}{1-(1-\pi)^n}\\
     & = 1- n\pi \frac{(1-\pi)^{n-1}}{1-(1-\pi)^n}\\
      & = \frac{n-1}{2}\pi + \frac{ 6n- {n^2}  -{5}}{12}\pi^2 +  o(n^2 \pi^2) \quad  \text{ \footnotesize (Taylor Expansion around $\pi=0$)}\\
            & \leq  \frac{n}{2}\pi \\
        \end{split}\label{eq:eq}
\end{equation}
Thus (i) $\pi \to \mathbb{P}[S \geq 2| S \geq 1 ] $ is an increasing function of $\pi$, so the higher the community prevalence, the higher the mode of the number of positive samples in infected pools, and (ii), in the regime that we are considering  $n\pi \leq 0.10$, so that $    \mathbb{P}[S \geq 2| S \geq 1 ]  \leq 0.05$. Thus, with 95\% confidence, we have only one contaminated sample per positive pool.

\underline{$H_a$: With the effect of correlations}:  $S = \sum_{i=1}^n T_i^{\text{(community)}} + \tilde{S} $ where $\tilde{S}$ denotes the sum of all network transmissions. 
$\tilde{S}| S \sim \text{Binom}(n-S, 1-(1-\tau)^S)$
Thus, the number of people infected due to the network is:
\begin{equation*}
    \begin{split}
        \mathbb{E}[S] &= n\pi + \sum_{k=1}^n { n \choose k} \pi^k (1-\pi)^{n-k} (n-k) (1-(1-\tau)^k)\\
        &= n\pi + (1-\pi)n (1- (1-\tau\pi)^{n-1})\\
        &= (1-\pi)n (1- 1 + (n-1)\tau\pi)  + o(n\pi) \\
         &= n\pi + n\pi (1-\pi)^{n-1} (n-1)\tau  + o(n\pi)    \\
              &= n\pi (1 +  (n-1)\tau ) + o(n\pi)  \\
    \end{split}
\end{equation*}
More interestingly, conditionally on considering infected pools, the probability that there are more than 1 infected samples in the pool is:
\begin{equation}
    \begin{split}
        \mathbb{P}_{H_a}[\sum_{i=1}^n Y_i > 1] &= 1- (1-\pi)^{n} - n\pi(1-\pi)^{n-1} (1-\tau)^{n-1} \\
        &= 1 - (1-\pi)^{n-1} ( 1 - \pi + n \pi (1-\tau)^{n-1}) \\
        &\geq  1 -  (1-(n-1)\pi + \frac{(n-1)(n-2)}{2}\pi^2  + o(n^2\pi^2) )\\
        &\times  ( 1 -\pi + n \pi (1-\tau)^{n-1} )\\
        &\geq  n\pi ( 1 - (1-\tau)^{n-1})   +  \pi^2 (n-1) ( \frac{n-4}{2} + n(1-\tau)^{n-1})\\
        &\geq n\pi ( 1 - (1-\tau)^{n-1}) 
    \end{split}\label{eq:proba_ha}
\end{equation}

We also have, by Taylor expansion around $\pi=0$:
\begin{equation*}
    \begin{split}
        \mathbb{E}[S | S>0] &= 1 + \sum_{k=1}^n { n \choose k} \frac{\pi^k (1-\pi)^{n-k}}{1-(1-\pi)^n} (n-k) (1-(1-\tau)^k)\\
        &= \frac{n\pi + (1-\pi)n (1- (1-\tau\pi)^{n-1})}{1-(1-\pi)^n}\\
              &\approx  1 + (n-1)\tau   +\frac{1}{2} (n-1)(1-\tau) ( (n-2)\tau+1) \pi + O(n^3\pi^2) \\
    \end{split}
\end{equation*}

where the remainder is bounded by $n^3\pi^2$. Since we have $n^3\pi^2 \leq \frac{n}{100}$, for the pool sizes we are considering (less than 50),  $ \mathbb{E}[S | S>0]$ is dominated by the second order polynomial $1 + (n-1)\tau   +\frac{1}{2} (n-1)(1-\tau) ( (n-2)\tau+1) \pi$. 

This means that the mean increases by $k$ as soon as $\tau$ is above the root of this polynomial

\pagebreak
\section{Simulations}\label{sec:appendix_sim}
\subsection{Mathematical Formulations of Probability Laws}
In computing the metrics of interest, we must also compute $p_k = P[\sum_{i=1}^n Y_i=K]$, the probability of $K$ positives in a pool of size $n$. The probability law for computing this depends on whether we invoke the null or alternative (network transmission) models, i.e. whether we account for correlation between members of a pool.  In the null model, the probability of observing $K$ positives in a pool of size $n$ is given by the binomial distribution: $p_k \sim Binom(K,n)$. When independence assumptions are violated (correlated individuals), we calculate the probability of having $K$ total positives in a pool of size $n$, $k$ of which are infected in the community (where probability of infection equals prevalence, $\pi$) and $K-k$ of which are infected via network transmission (where probability of infection equals a homogeneous network transmission probability between all individuals, $\tau$). We calculate the total probability over all possible values of $k$ (Equation \ref{eq:positive_probability}). 

\begin{equation}   
    \begin{split}
     P(\sum Y_i = K)  
     &= \sum_{k=1}^K \left( {n \choose k} \pi^k (1-\pi)^{n-k} {n-k \choose K- k} (1-(1-\tau)^k)^{K-k} ((1-\tau)^k)^{(n-K)} \right)
     \end{split} \label{eq:positive_probability}
\end{equation}

If we allow $\tau$ to be heterogeneous between groups, but homogeneous within a group, (i.e. each groups has its own $\tau_i$) and then consider the probability of seeing $K$ positives average over the total number of groups ($n_{group}$), we obtain Equation \ref{eq:positive_prob_tauhet}:
\begin{equation}   
    \begin{split}
     P(\sum Y_i = K) = 
     \sum_{k=1}^K \left( {n \choose k} \pi^k (1-\pi)^{n-k} \sum_{\tau_i=1}^{n_{groups}} P(\tau = \tau_i) {n-k \choose K- k} (1-(1-\tau_i)^k)^{K-k} ((1-\tau_i)^k)^{(n-K)} \right)
     \end{split} \label{eq:positive_prob_tauhet}
\end{equation}
where $P(\tau = \tau_i) = \frac{1}{n_{groups}}$

Similarly, we can further account for heterogeneity in risk among individuals by allowing $\pi$ to vary, reflecting the fact that individuals may have greater or less risk of being infected from the community depending on their behaviors, profession, etc. (Equation \ref{eq:positive_prob_pihet}).
\begin{equation}   
    \begin{split}
     P(\sum Y_i = K) = 
     \sum_{k=1}^K \left( \sum_{pi_i=1}^{n_{groups}} P(\pi = \pi_i) {n \choose k} \pi_i^k (1-\pi_i)^{n-k} {n-k \choose K- k} (1-(1-\tau_i)^k)^{K-k} ((1-\tau_i)^k)^{(n-K)} \right)
     \end{split} \label{eq:positive_prob_pihet}
\end{equation}
where $P(\pi = \pi_i) = \frac{1}{n_{groups}}$

Combining equations \ref{eq:positive_prob_tauhet} and \ref{eq:positive_prob_pihet} represents the model in which we account for heterogeneity in both $\tau$ and $\pi$.

\subsection{$C_t$ Value Data}
We use empirically collected data on the distribution of $C_t$ values from \cite{Wang2021}. Briefly, nasopharyngeal or oropharyngeal swab specimens obtained for SARS-CoV-2 testing were obtained by the Stanford Clinical Virology Laboratory from tertiary-care academic hospitals and affiliated outpatient facilities in the San Francisco Bay Area, California, from June 10 - June 19, 2020 and July 6 - July 23, 2020. Samples were collected both from symptomatic and asymptomatic inpatients and outpatients, either for clinical care or via COVID-related surveillance studies and drug trials.  

The distribution of $C_t$ values is best represented by a Weibull distribution with shape parameter $s=4.5$ and scale $\eta = 30$. The distribution of the $C_t$ values depends on a number of factors, including the population tested (ie., hospital admissions vs general population, COVID variant, etc). To create a realistic distribution of $C_t$ values with the appropriate amount of spread, we sample and shift the Weibull distribution of \citeauthor{Wang2021}: we sample from their fitted distribution to create a mock distribution of individual $C_t$ values, and shift it so that 25.0\% of samples are above $C_t= 35$. 

\subsection{Prior Distributions on $\pi$ and $\tau$}
To sample $\pi$ and $\tau$ from informed priors, we fit Beta distributions to published data on prevalence and SAR, respectively. Beta distributions to the 95\% confidence intervals of reported metrics of interest (prevalence, SAR) using the publicly available \texttt{beta.params.from.quantiles} function.\footnote{\url{http://www.medicine.mcgill.ca/epidemiology/Joseph/PBelisle/BetaParmsFromQuantiles.html}}

\begin{figure}[H]
    \includegraphics[width=\linewidth]{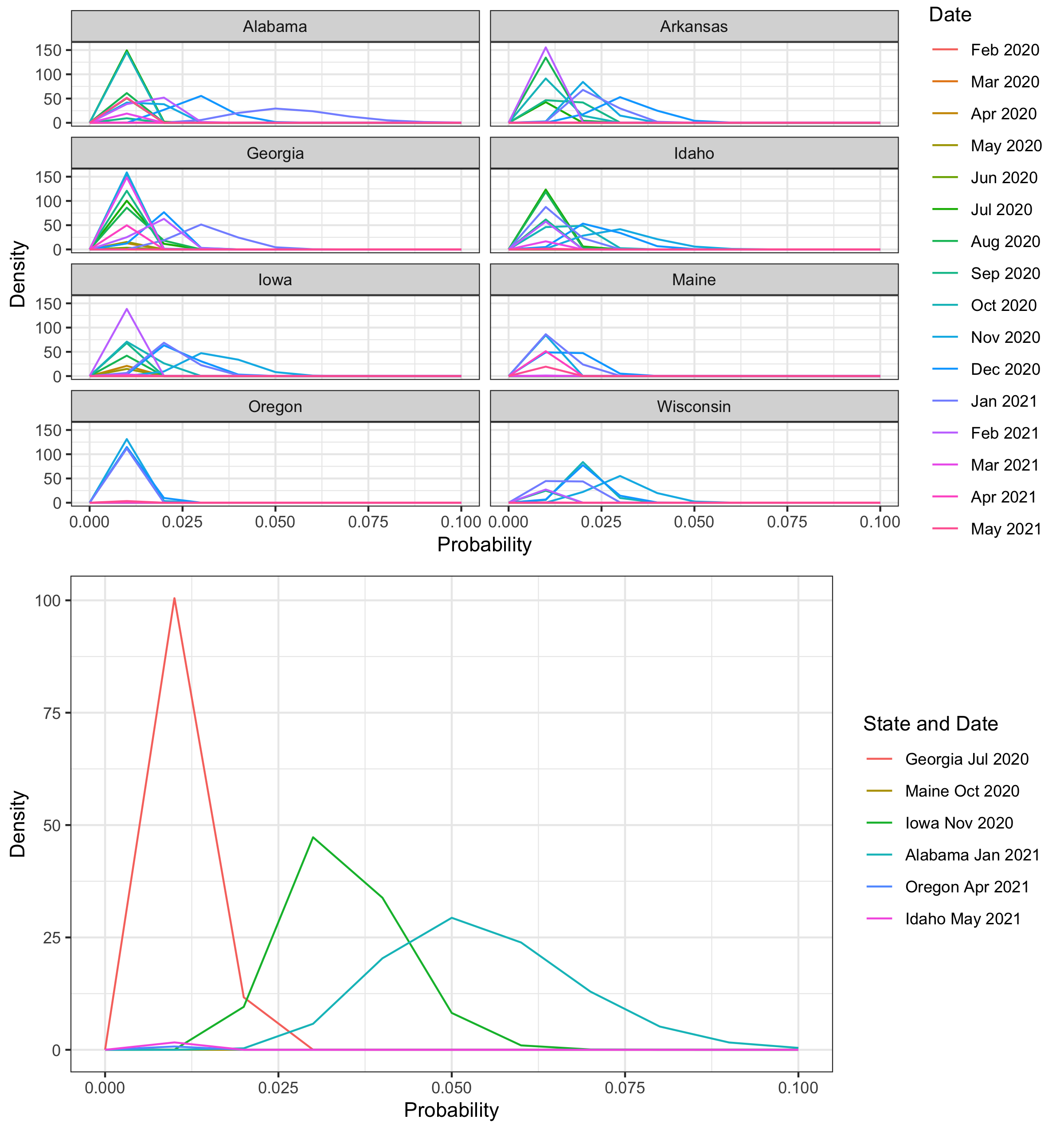}\hfill
  \caption{Density plots of the Beta prior distributions for $\pi$, corresponding to all fitted models (top) and the parameters and settings selected for the simulation study and included in Table \ref{table:pi_dist}}
   \label{fig:pi_distributions}
\end{figure}

To estimate distributions for the community prevalence, we fit a Beta distribution to reported 95\% confidence intervals on the estimated rates of COVID-19 infections over time in every U.S. state. Specifically, we use the estimate of true number of infections, which is adjusted for reporting delays and potential under-counting, provided by \url{https://covidestim.org/}. The methodology for adjusting case counts is described in a pre-print paper by the authors.\footnote{\url{https://www.medrxiv.org/content/10.1101/2020.06.17.20133983v2}} 
Data extend from the first reported case (January 13, 2020) through present (data downloaded 28 May 2021). We only consider states for which 95\% CI data are available, and then further subset to consider a single representative state from each geographic division as defined by the U.S. census bureau. Prior to fitting the distribution, we normalize raw infection counts by the state population for 2019 \footnote{\url{https://www.census.gov/data/tables/time-series/demo/popest/2010s-state-total.html}} and sum over the past ten days of data to estimate active cases. Finally, we average the number of active cases by month for each state. After fitting Beta distributions to the resulting data, we observed that many prior distributions have very similar $\alpha, \beta$ parameters. Thus, we select six representative distributions corresponding to different stages in the pandemic (e.g. large surge, small surge, peak of surge, low cases, declining cases), and varying time points and geographic regions. The selected time points and regions, as well as corresponding fitted distributions, are presented in Table~\ref{table:pi_dist}; density plots for all fitted distributions and those selected for use in the simulations are presented in Figure~\ref{fig:pi_distributions}.

\begin{figure}[H]
    \includegraphics[width=\linewidth]{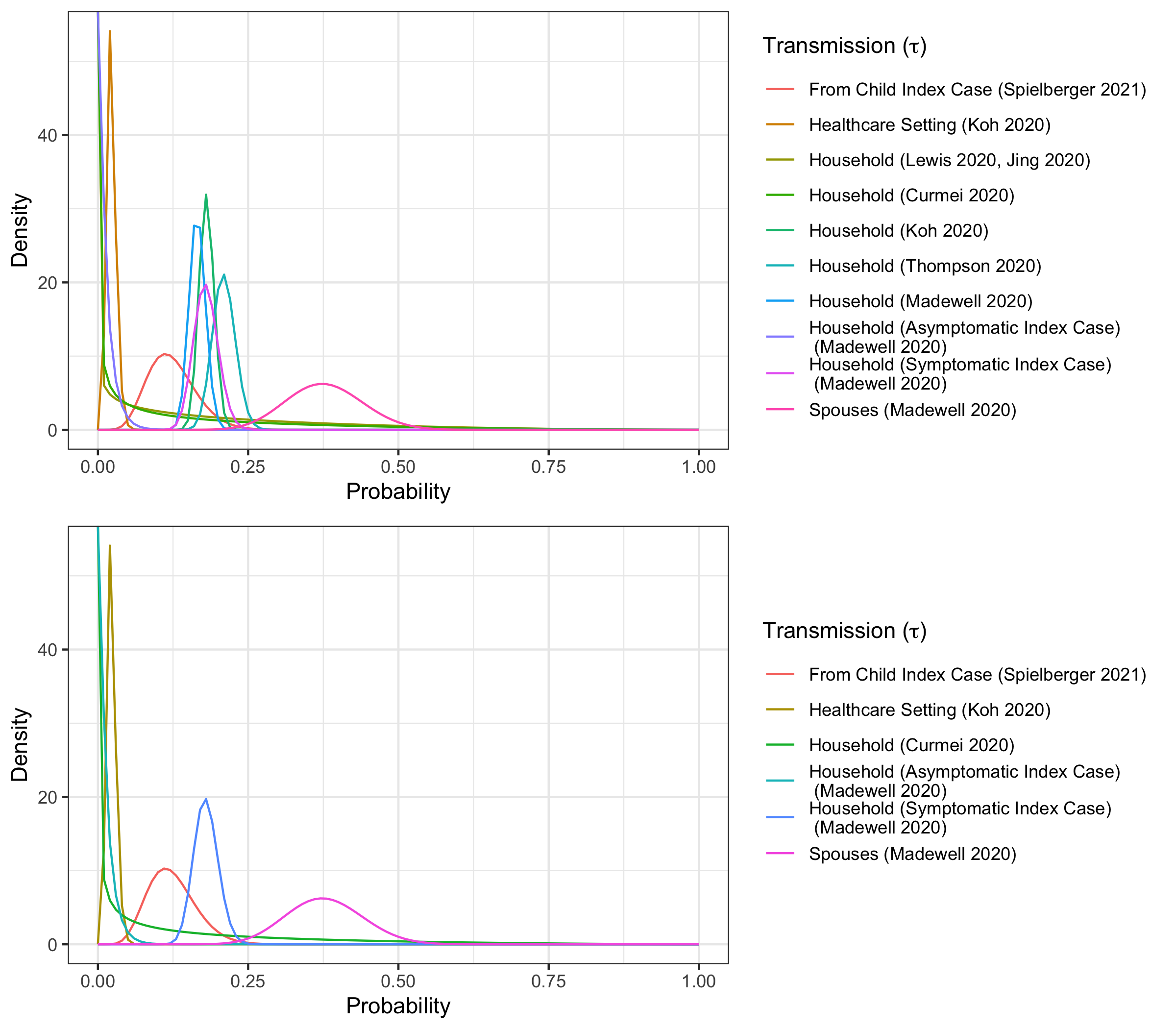}\hfill
  \caption{Density plots of the Beta prior distributions for $\tau$, corresponding to all fitted models (top) and the parameters and settings selected for the simulation study and included in Table \ref{table:tau_dist}}
   \label{fig:tau_distributions}
\end{figure}

To estimate distributions for network transmission, we followed a similar procedure as described above for community prevalence. Beta distributions were fit to the 95\% confidence intervals of SAR estimates reported in published meta-analyses for different settings. Again, many fitted distributions were similarly specified and so six representative distributions were selected (Figure~\ref{fig:tau_distributions}). The settings and corresponding parameters of the selected distributions are presented in Table~\ref{table:tau_dist}; density plots of all fitted distributions and the representative selected distributions for use in the simulations are presented in Figure~\ref{fig:tau_distributions}. 

\begin{table}[h]
\caption{Parameters for select Beta prior distributions for community prevalence $\pi$ in different settings. The Beta parameters ($\alpha, \beta$) are estimated from the 95\% confidence intervals of the mean estimated active cases in a given state and month, obtained from covidestim.org.} \label{table:pi_dist}
\resizebox{\textwidth}{!}{
\begin{tabular}{ ccc } 
 \hline
Setting & Mean Monthly Prevalence (95\% CI) & Beta$(\alpha, \beta)$ \\
 \hline
 Georgia, July 2020 &  1.3\% (0.7, 2.0) & Beta(16.67, 1282.88) \\
 Maine, October 2020 & 0.2\% (0.07, 0.3) & Beta(9.94, 6561.33) \\
 Iowa, November 2020 & 3.4\% (2.0, 5.2) & Beta(16.99, 477.12)\\
 Alabama, January 2021 & 5.4\% (3.0, 8.4) & Beta(14.38, 251.01)\\
 Oregon, April 2021 & 0.5\% (0.2, 0.7) & Beta(13.06, 2836.41)\\
 Idaho, May 2021 & 0.4\% (0.1, 0.7) & Beta(5.77, 1543.33)\\
 \hline
\end{tabular}
}
\end{table}

\begin{table}
\caption{Parameters for select Beta prior distributions for network transmission $\tau$ in select settings. The Beta parameters ($\alpha, \beta$) are estimated from the 95\% confidence intervals of secondary attack rate (SAR) values in the literature. } \label{table:tau_dist}
\resizebox{\textwidth}{!}{
\begin{tabular}{ cccc } 
 \hline
Setting & SAR (95\% CI)  & Beta$(\alpha, \beta)$ & Citation \\
 \hline
 Child Index Case & 13.40\% (5.7-21.1) & Beta(8.38, 59.43) & Spielberger et al 2021\\
 Healthcare Setting & 0.7\% (0.4-1.0) & Beta(8.3, 359.61) & Koh et al 2020\\
 Household (Spouses) & 37.8\% (25.8-50.5) & Beta(21.78, 35.92) & Madewell et al 2020\\
 Household (Asymptomatic Index Case) & 0.7\% (0-4.9) & Beta(0.74, 62.23) & Madewell et al 2020\\
 Household (Symptomatic Index Case) & 18.0\% (14.2-22.1) & Beta(64.95, 296.26) & Madewell et al 2020\\
 Household (General) & 30\% (0-67) & Beta(0.45, 2.37) & Curmei et al 2020\\
 \hline
\end{tabular}
}
\end{table}

\subsection{Implementing Simulations}
We perform simulations under four distinct settings:
\begin{itemize}
    \item Fixed (point estimates of $\pi$ and $\tau$): Deterministic $\pi$ and $\tau$, probability calculated as in Equation \ref{eq:positive_probability}
    \item $\tau$ Graph Effect (Random $\tau$, Fixed $\pi$): Sample $\tau$ from a Beta prior, probability calculated as in Equation \ref{eq:positive_prob_tauhet}
    \item $\pi$ Graph Effect (Random $\pi$, Fixed $\tau$): Sample $\pi$ from a Beta prior, probability calculated as in Equation \ref{eq:positive_prob_pihet}
    \item All ($\tau$ and $\pi$) Graph Effect (Random $\tau$, Random $\pi$): Sample both $\pi$ and $\tau$, combining Equations \ref{eq:positive_prob_tauhet} and \ref{eq:positive_prob_pihet}.
\end{itemize}

\begin{figure}[H]
    \includegraphics[width=\linewidth]{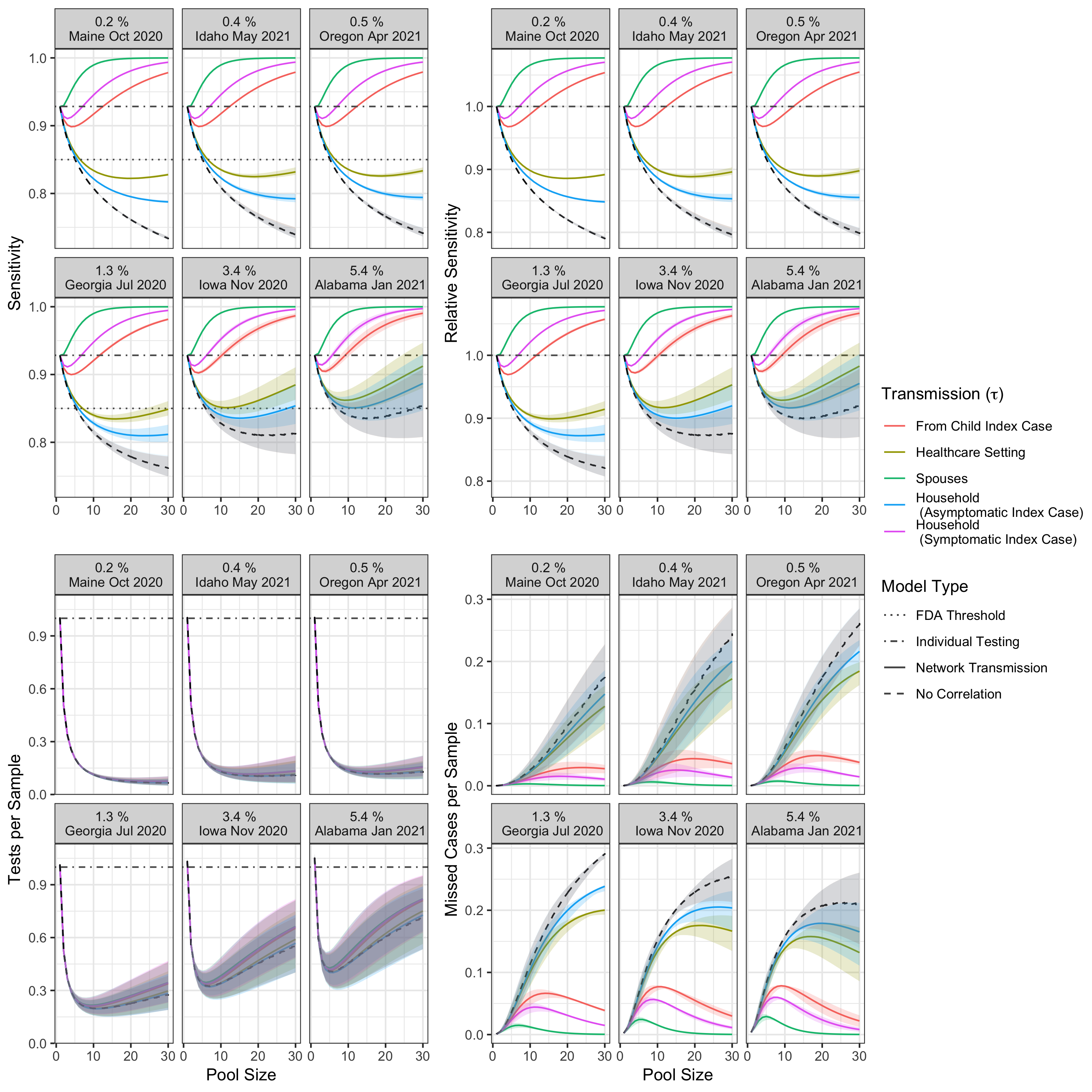}\hfill
  \caption{\textbf{Random Prevalence/Fixed Network Effect Model (Pi Graph Effect)}: Model-estimated parameters (sensitivity, relative sensitivity, expected tests per sample, missed cases per sample) by pool size, prevalence ($\pi$), and network transmission probability ($\tau$) for the Pi Graph Effect model, where $\pi$ is sampled from Beta distributions corresponding to the specified scenario. $\tau$ is held constant at the mean of the Beta distribution for the specified scenario. Shaded area indicates the empirical 95 percent credible prediction interval. Proportion of samples with $C_t$ above the 95\% LoD (held constant at LoD = 35) is constant, equal to 25\%. The null (no correlation) model, individual testing, and FDA sensitivity threshold are also indicated, where relevant.}
   \label{fig:allmetrics_pigraph}
\end{figure}

Results for All Graph Effect and $\tau$ Graph Effect are presented in the main paper; results for the Fixed model and $\pi$ Graph Effect are presented in Figures \ref{fig:allmetrics_fixed} and \ref{fig:allmetrics_pigraph}.

 \begin{figure}[H]
    \includegraphics[width=\linewidth]{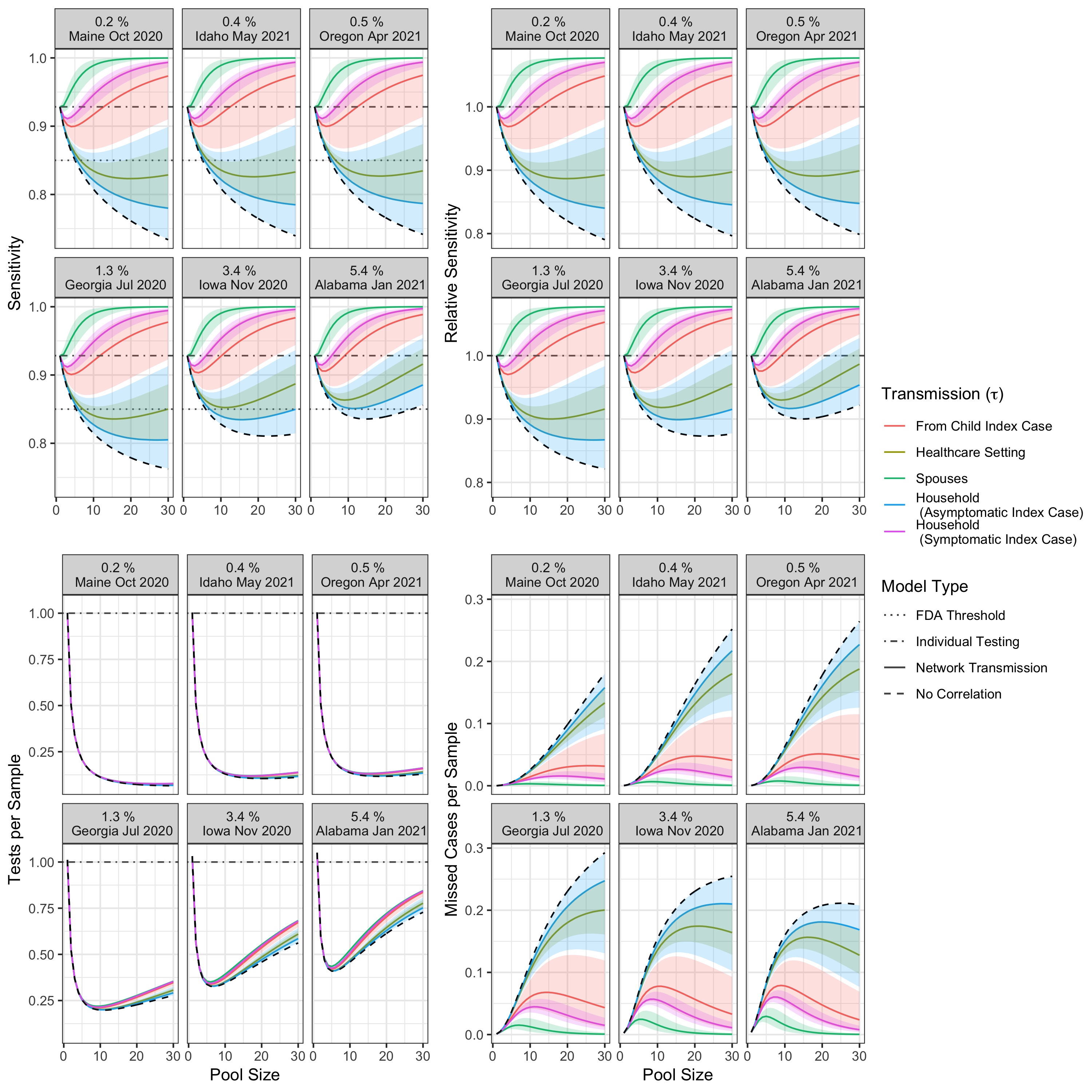}\hfill
  \caption{\textbf{Random Network Effect/Fixed Prevalence Model (Tau Graph Effect)}: Model-estimated parameters (sensitivity, relative sensitivity, expected tests per sample, missed cases per sample) by pool size, prevalence ($\pi$), and network transmission probability ($\tau$) for the Tau Graph Effect model, where $\tau$ is sampled from Beta distributions corresponding to the specified scenario. $\pi$ is held constant at the mean of the Beta distribution for the specified scenario. Shaded area indicates the empirical 95 percent credible prediction interval. Proportion of samples with $C_t$ above the 95\% LoD (held constant at LoD = 35) is constant, equal to 25\%. The null (no correlation) model, individual testing, and FDA sensitivity threshold are also indicated, where relevant.}
   \label{fig:allmetrics_taugraph}
\end{figure}

\pagebreak
\section{Complementary material}\label{sec:appendix_complementary}

\begin{figure}[h]
    \centering
    \includegraphics[width=\textwidth]{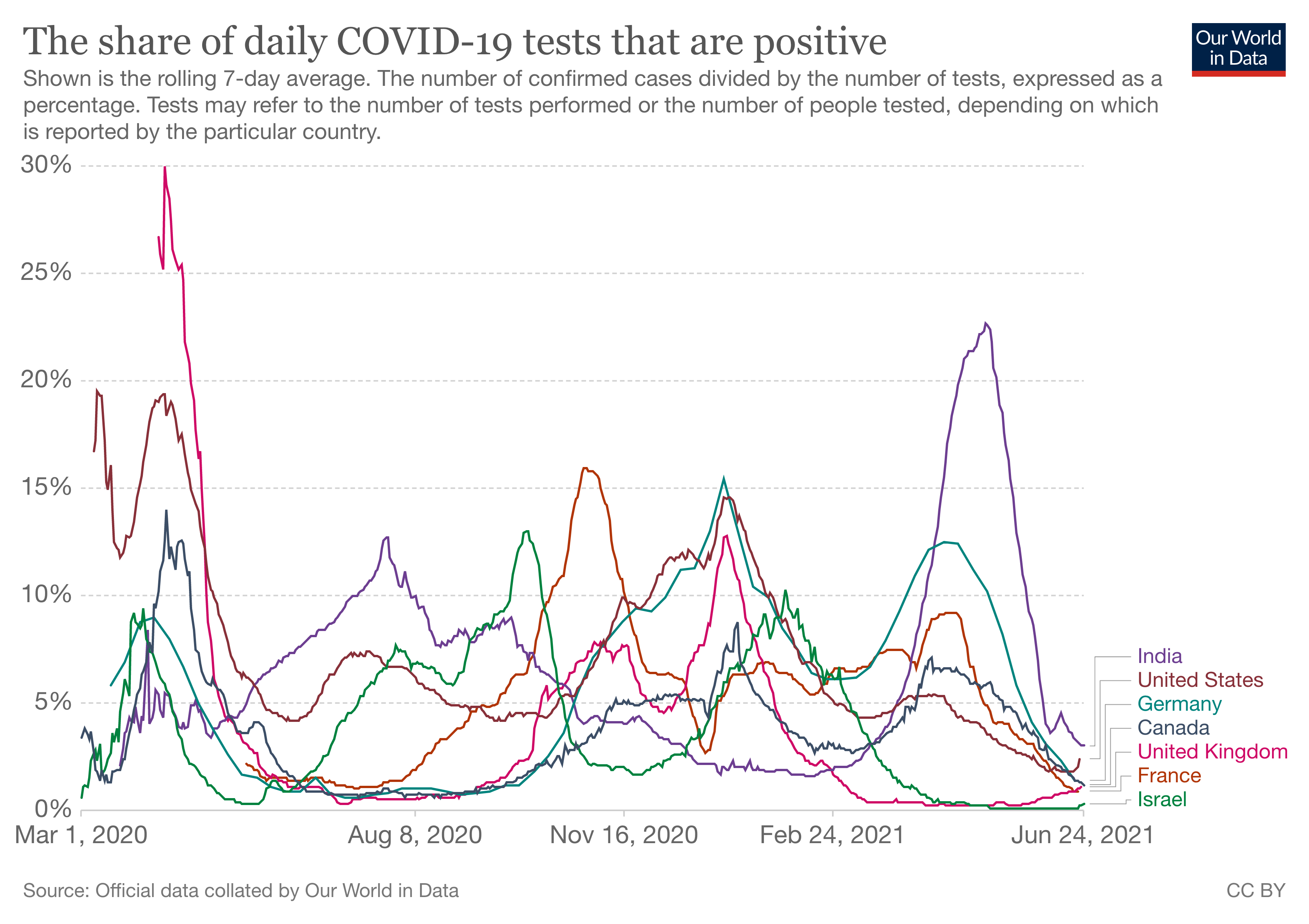}
    \caption{Chart from \href{https://ourworldindata.org/covid-cases}{Our World in Data} showing the prevalence levels among tests in different countries throughout the pandemic. }
    \label{fig:comp}
\end{figure}

\end{document}